\documentclass[times,preprint]{elsarticle}

\usepackage{cnf}

\usepackage{framed,multirow}

\usepackage{amssymb}
\usepackage{latexsym}

\usepackage{url}
\usepackage{xcolor}
\definecolor{newcolor}{rgb}{.8,.349,.1}
\usepackage[acronym]{glossaries}
\usepackage{hyperref}
\usepackage{booktabs}
\usepackage{caption}
\usepackage{makecell}
\usepackage{ulem}

\makeglossaries

\newacronym{les}{LES}{Large-Eddy Simulations}
\newacronym{dns}{DNS}{Direct Numerical Simulations}
\newacronym{jpdf}{jPDF}{joint-Probability Density Function}
\newacronym{pdf}{PDF}{Probability Density Function}
\newacronym{asgs}{ASGS}{Algebraic Subgrid Scale}
\newacronym{tc}{TC}{Tabulated chemistry}

\definecolor{salmon}{HTML}{FA8072}

\newcommand{\ef}[1]{{\color{black}{#1}}}

\newcommand{\dm}[1]{{\color{black}{#1}}}

\usepackage[switch,pagewise]{lineno} 

\journal{Combustion \& Flame}

\begin{document}

\verso{Emiliano M. Fortes et al.}

\begin{frontmatter}

\title{Large-eddy simulations of a lean hydrogen premixed turbulent jet flame with tabulated chemistry\tnoteref{tnote1}}%

\author[1]{Emiliano M. \snm{Fortes}}
\cortext[cor1]{Barcelona Supercomputing Center (BSC), Plaça Eusebi Güell, 1-3 08034, Barcelona, Spain}

\author[1]{Eduardo J. \snm{Pérez-Sánchez}}
\author[2]{Temistocle \snm{Grenga}}
\author[3]{Michael \snm{Gauding}}
\author[3]{Heinz \snm{Pitsch}}
\author[1]{Daniel \snm{Mira} \corref{cor1}}
\emailauthor{daniel.mira@bsc.es}{Daniel Mira}


\address[1]{Barcelona Supercomputing Center (BSC), Plaça Eusebi Güell, 1-3 08034, Barcelona, Spain}
\address[2]{Faculty of Engineering and Physical Sciences, University of Southampton, SO17 1BJ, Southampton, United Kingdom}
\address[3]{Institute for Combustion Technology, RWTH Aachen University, 52056 Aachen, Germany}

\begin{abstract}
Large-eddy simulations (LES) of a planar turbulent lean hydrogen–air jet flame at $Re = 11000$ are performed using a tabulated flamelet model based on mixture-averaged diffusion that incorporates detailed transport, including differential and preferential diffusion, wall heat loss, and thermodiffusion. The approach is extended to turbulent combustion in LES using a presumed-shape probability density function formulation that accounts for sub-filter effects. The flame exhibits a highly corrugated front, driven by local variations of mixture fraction induced by strong thermodiffusive transport.
These effects significantly alter both the flame structure and morphology.
The LES results are systematically compared to a reference direct numerical simulation across varying \ef{LES filters through different} mesh resolutions to evaluate the predictive capability of the model. The LES accurately reproduces instantaneous flow structures and thermodiffusive effects. Global flame characteristics including flame length, surface area, and consumption speed, are well captured and show limited sensitivity to mesh resolution.
The role of thermodiffusion is also examined, showing that its incorporation leads to a more reactive flame and should not be neglected in the formulation. Heat losses are incorporated into the tabulated chemistry framework for completeness but are found to have a negligible impact, consistent with the walls weak influence in the present configuration.
Overall, the results demonstrate that the proposed approach provides reliable predictions of the main flame characteristics, with remaining discrepancies primarily associated with unresolved sub-filter effects that deserve further investigation.
\end{abstract}


\end{frontmatter}


\section*{Novelty and significance statement}
This work introduces a significant advance in the predictive modelling of lean hydrogen turbulent flames by formulating a tabulated chemistry framework that captures the full complexity of mixture-averaged transport, preferential and differential diffusion, wall heat loss, and thermodiffusion in the context of LES.
Unlike prior studies that rely on simplified transport closures or constant Lewis numbers, the present approach embeds detailed molecular and thermal diffusion physics while retaining computational efficiency. Its significance lies in demonstrating that such a comprehensive treatment of transport phenomena enables accurate representation of thermodiffusive behaviour and flame topology in turbulent hydrogen combustion. This establishes a more physically grounded basis for modelling low-Lewis-number flames and strengthens the reliability of flamelet-based LES for hydrogen combustion.

\section{Introduction}

Hydrogen has emerged as a leading candidate towards net-zero CO$_2$ emissions due to its strong potential to drive sustainable energy solutions. However, several challenges are associated with the use of this fuel, including high flame speed, high mass diffusivity, low volumetric energy density, NO$_\mathrm{x}$ emissions, and enhanced instabilities, all of which add complexity to the deployment of hydrogen for industrial applications~\cite{PITSCH2024105638}.
Burning hydrogen under lean conditions offers a practical mitigation strategy, as it moderates some of these adverse effects, particularly NO$_\mathrm{x}$ formation. 
However, this benefit comes at the cost of enhanced thermodiffusive effects, whose accurate representation remains a modelling challenge for \gls{les}~\cite{BergerUpcoming}.

In hydrogen flames, differential and preferential transport play an important role within the flame structure. \ef{Preferential} diffusion arises from the disparity between mass and thermal diffusivities, leading in lean hydrogen–air mixtures to a very low fuel Lewis number ($Le_{\mathrm{H_2}} \approx 0.3 \ll 1$). Preferential diffusion results from unequal species diffusivities, such that $Le_i$ varies among species. While present in all fuels, these effects are particularly pronounced in lean hydrogen flames, where they strongly influence the burning velocity and promote thermodiffusive instabilities~\cite{Berger2022_partI,Fortes2024,Matalon2018}. Detailed analyses identify differential diffusion as the primary mechanism driving flame corrugation~\cite{PITSCH2024105638} and, consequently, exerting a dominant influence on global flame characteristics.  
Differential diffusion contributes to local enrichment in flame regions of positive curvature or fuel depletion in negative curvature regions (seen from fresh gases), further promoting the formation of thermodiffusive instabilities \cite{PITSCH2024105638}. In addition, the Soret effect further enhances these instabilities, although its contribution is comparatively weaker \cite{GRCAR20091173,Zhou2017EffectOfSoretDiffusion,Schlup2017ValidationOfAMixtureAveraged}.

\gls{tc} methods offer a powerful strategy for simulating turbulent combustion, providing a practical route to incorporate complex transport and chemical phenomena while maintaining computational efficiency~\cite{Mira2023HPCenablingTechnologiesHighfidelity}.
Accordingly, a range of modelling strategies has been developed to account for these thermodiffusive effects within \gls{tc} frameworks.
Early approaches focused on determining effective Lewis numbers from 1D laminar structures~\cite{Vreman2009} and on introducing additional transport terms and modulated flux coefficients into the governing equations for the controlling variables~\cite {deSwart2010,Donini2015,Abtahizadeh2015,Nicolai2022}. This methodology has been successfully applied to stratified hydrogen flames, blends, and \gls{les} configurations~\cite{Dinesh2013,Donini2017,Almutairi2023}. Alternative formulations have sought to improve accuracy by reformulating the transport terms. Regele et al.~\cite{Regele2013} and Schlup et al.~\cite{Schlup2019} introduced source terms to account for differential diffusion and the Soret effect within the mixture fraction equation. Other strategies involve transporting species mass fractions to reconstruct the manifold coordinates~\cite{Bottler2022,Bottler2023} or using commuting operators under constant Lewis number assumptions to derive new transport coefficients~\cite{Mukundakumar2021}. 
While these models have demonstrated excellent predictive capabilities in laminar flame configurations, their application to turbulent flames subjected to strong thermodiffusive effects is less common.
Nicolai et al.~\cite{Nicolai2022} applied a high-dimensional tabulation strategy to explicitly account for differential diffusion in stratified hydrogen–methane flames using \gls{les} and quantified its impact with experimental data. Almutairi et al.~\cite{Almutairi2023} used a \gls{tc} model including preferential diffusion for Reynolds-Averaged Navier-Stokes simulations of hydrogen-enriched flames on a dual-fuel engine, showing improved agreement with experiments in terms of pressure traces and heat-release rates compared with diffusion-neutral tabulations.

More recently, several studies have focused specifically on lean hydrogen flames with thermodiffusive instabilities. Berger et al.~\cite{berger2022development} applied an \gls{les} combustion model incorporating the extended version of Regele's model~\cite{Regele2013} for thermodiffusive transport by Schlup and Blanquart \cite{Schlup2017ValidationOfAMixtureAveraged,Schlup2019} into a filtered progress variable/mixture fraction description, and performed \textit{a priori} and \textit{a posteriori} assessments using \gls{dns} data~\cite{Berger2022_dns}. Ferrante et al.~\cite{Ferrante2024} investigated different closures for differential diffusion, including the models of Regele et al.~\cite{Regele2013} and Mukundakumar et al.~\cite{Mukundakumar2021} for partially premixed hydrogen flames in the framework of presumed ﬁltered probability density function for \gls{les}.
Masucci et al.~\cite{Masucci2025NumericalInvestiongOfPremixed} assessed the capabilities of \gls{les} combined with flamelet models for both premixed and non-premixed hydrogen flames, highlighting the potential of tabulated approaches for low-emission gas-turbine applications, but also pointing out limitations related to preferential diffusion and near-wall phenomena. 

Despite these advances, validations of tabulated chemistry models against \gls{dns} for turbulent hydrogen flames subject to strong thermodiffusive instabilities remain limited. The objective of the present work is to address this gap through a comprehensive assessment of a \gls{tc} framework based on mixture-averaged diffusion developed in previous studies by the authors~\cite{PerezSanchez2023}, which has recently been validated in several studies \cite{Fortes2024,BOTTARI2026136805,Schneider_Rong_Hasse_Nicolai_2026}, but extended to LES using a presumed-shape PDF approach~\cite{Donini2017,Govert2018}. Particular attention is given to evaluating the sensitivity of the predictions to mesh resolution and to the treatment of complex transport phenomena. To this end, \gls{les} results are benchmarked against the high-fidelity \gls{dns} data of a planar turbulent premixed hydrogen–air jet flame investigated by Berger et al.~\cite{Berger2022_dns}. 
In addition to the grid-resolution study, a set of simulations are defined to isolate the effects of wall heat loss and thermodiffusion, enabling a quantitative assessment of their relative impact on flame topology, consumption speed, flame length, and overall model accuracy.

The paper is organised as follows. Section~\ref{sec:model_descr} introduces the tabulated-chemistry formulation, including its extension to LES. Section~\ref{sec:numerical_fwrk} details the numerical framework, including the transport equations for \gls{les}, sub-filter models, and the construction of the laminar flame manifolds. Section~\ref{sec:case_config} describes the planar turbulent premixed jet configuration, the reference \gls{dns}, and the set of LES cases with the different physics comprised. The results are presented in Section~\ref{sec:results}, where a dedicated analysis is conducted to assess the predictive capabilities of the different modelling features across different mesh resolutions.
Finally, conclusions and perspectives for future work are summarised in Section~\ref{sec:conclusions}.

\section{Model description}\label{sec:model_descr}

The modelling of hydrogen flames requires the accurate prediction of diffusive fluxes across the flame front to account for preferential and differential diffusion. To capture these effects, species diffusion is modelled using a  mixture-averaged diffusion formulation, which accounts for the effect of the mixture molecular weight and a velocity correction for mass conservation \cite{Giovangigli1991ConvergentIterativeMethods,PerezSanchez2023}.
Additionally, thermal diffusion, also known as the Soret effect, is included due to its significance when the flame is curved \cite{Schlup2019}.

For a mixture of $N_s$ species the mixture-averaged diffusion model \ef{with the Soret effect} approximates the mass diffusion fluxes for each species $k=1,\ldots,N_s$ as:
\begin{equation}\label{eq:diff_eq_MA}
\mathbf{V}_k Y_k = - \frac{W_k}{W} D_k \nabla X_k - \frac{D_k^T}{\rho T} \nabla T,
\end{equation}

\noindent where $\mathbf{V}_k$ is the diffusion velocity of the $k$-th species (vectors are denoted in bold characters), $X_k$ and $Y_k$ are its molar and mass fractions, respectively, $D_k = 1-Y_k / \sum_{\substack{j=1,j \neq k}}^{N_s} X_j / \mathcal{D}_{jk}$ is the diffusion coefficient, $W_k$ and $W$ denote the molecular weights of the species and the mixture, respectively, $\mathcal{D}_{jk}$ are the binary diffusion coefficients of the $j$-th into the $k$-th species, \ef{$T$ denotes the temperature and $D_k^T$ is the thermal diffusion coefficient of the $k$-th species.}

\ef{Rewriting the species flux in terms of the mass fraction and including the velocity correction $\mathbf{V_c}$ (such that $\sum_{j=1}^{N_s} (\mathbf{V}_k+\mathbf{V_c})Y_k=0$), the mass diffusive flux for the $k$-th species takes the form:}
\begin{align}\label{eq:diff_eq_MA_total}
\mathbf{j}_k = {} & - \frac{W_k}{W} D_k \nabla X_k
+ Y_k \sum_{j=1}^{N_s} \frac{W_j}{W} D_j \nabla X_j \notag \\
& - \frac{D_k^T}{\rho T} \nabla T
+ Y_k \sum_{j=1}^{N_s} \frac{D_j^T}{\rho T} \nabla T .
\end{align}

In \gls{tc} methods, the manifold representing all possible thermochemical states $\mathbf{\Psi}$ of the flames is parametrised by a set of $N_c$ controlling variables $(\mathcal{Y}_1, \ldots, \mathcal{Y}_{N_c})$, such that $\mathbf{\Psi}=\mathbf{\Psi}(\mathcal{Y}_1, \ldots, \mathcal{Y}_{N_c})$. Therefore, the physical gradient of the dependent variables can be expressed through the chain rule in terms of the gradients of the controlling variables and the gradients of the manifold’s dependent variables in phase space. Accordingly, the gradient operator in the \gls{tc} framework can be rewritten as $\nabla = \sum_{i=1}^{N_c} \nabla \mathcal{Y}_i \frac{\partial}{\partial \mathcal{Y}i}$, and the species fluxes can be written in the compact form $\mathbf{j}_k = - \sum_{i=1}^{N_c} \Lambda_{Y_k,\mathcal{Y}_i} \nabla \mathcal{Y}_i$, where:
\begin{multline}\label{eq:chain_rule_flux}
\Lambda_{Y_k,\mathcal{Y}_i} = \frac{W_k}{W} D_k \frac{\partial X_k}{\partial \mathcal{Y}_i} 
- Y_k\sum_{j=1}^{N_s}D_j \frac{W_j}{W} \frac{\partial X_j}{\partial \mathcal{Y}_i} \\ + \frac{D_k^T}{\rho T}\frac{\partial T}{\partial \mathcal{Y}_i} - Y_k\sum_{j=1}^{N_s}\frac{D_j^T}{\rho T}\frac{\partial T}{\partial \mathcal{Y}_i}.
\end{multline}

\noindent The coefficients $\Lambda_{Y_k,\mathcal{Y}_i}$ depend only on the flame structure in phase space and, therefore, can be expressed as a function of $(\mathcal{Y}_1, \ldots, \mathcal{Y}_{N_c})$ and stored in look-up tables~\cite{PerezSanchez2023}.

If the fluxes of the control variables are defined as functions of the fluxes of the manifold-dependent variables, transport equations for the controlling variables can be derived. In particular, if they are defined as linear combinations of species mass fractions, the controlling variables transport equations can be derived by linearly combining the species transport equations, which are given by:
\begin{equation}\label{eq:transport_Yk} \rho \frac{\partial Y_k}{\partial t} + \rho \boldsymbol{u} \cdot \nabla Y_k + \nabla \cdot (\rho \,\mathbf{j}_k) = \dot{\omega}_k. \end{equation}

By applying the equality of Eq.~\eqref{eq:chain_rule_flux} \ef{to} Eq.~\eqref{eq:transport_Yk}, the controlling variables transport equations take the following general form:

\begin{equation}\label{eq:eq_Y_control_general}
\rho \frac{\partial \mathcal{Y}_i}{\partial t} + \rho \boldsymbol{u} \cdot \nabla \mathcal{Y}_i = \sum_{j=1}^{N_c} \nabla \cdot (\rho \, \Gamma_{\mathcal{Y}_i,\mathcal{Y}_j} \, \nabla \mathcal{Y}_j ) + \dot{\omega}_{\mathcal{Y}_i},
\end{equation}

\noindent where coefficients $\Gamma_{\mathcal{Y}_i,\mathcal{Y}_j}$ are the result of linearly combining coefficients $\Lambda_{Y_k,\mathcal{Y}_i}$ according to the definition of the controlling variables, and $\dot{\omega}_{\mathcal{Y}_i}$ denotes their associated source terms.
Using this notation, the diffusive flux for the controlling variables can be expressed as $\mathbf{j}_{\mathcal{Y}_i} = \sum_{j=1}^{N_c} (\Gamma_{\mathcal{Y}_i,\mathcal{Y}_j}\, \nabla \mathcal{Y}_j )$.

For the particular configuration studied in this work, wall heat losses are included in the injection \cite{Berger2022_dns}, and therefore, for completeness, the \gls{tc} method must also account for these effects. For manifolds with heat losses, the typical choice for the controlling variables includes a reactive progress variable $Y_c$, the mixture fraction $Z$ and the total enthalpy $h$. For the $h$ equation, the dissipation term due to viscous forces is neglected.
Then, the resulting transport equation \ef{has} the same structure \ef{as} Eq.~\eqref{eq:eq_Y_control_general} with no source term and a diffusive flux defined by Fourier's law, and given by: 
\begin{equation}\label{eq_jh}
 \mathbf{j_h} =-\sum_{i=1}^{N_c} \left(\frac{\lambda}{\rho} \frac{\partial T}{\partial \mathcal{Y}_i}+
 \sum_{k=1}^{N_s}\Lambda_{Y_k,\mathcal{Y}_i} h_k \right) \nabla \mathcal{Y}_i=- \sum_{i=1}^{N_c} \Gamma_{h,\mathcal{Y}_i} \nabla \mathcal{Y}_i,
\end{equation}

\noindent where $\lambda$ is the thermal conductivity and $h_k$ is the specific enthalpy for the $k$-th species~\cite{PerezSanchez2023}.

In practice, it is convenient to use normalised control variables for tabulation. Therefore, the space of $(Y_c,Z, h)$ is transformed into the manifold of $(c,Z, I)$ variables for the cases with heat loss, where $c$ and $I$ are the normalised progress variable and normalised enthalpy, respectively, defined as:

\begin{equation}\label{eq:normalization_Yc}
c(Y_c,Z)=\frac{Y_c-Y_{c,u}(Z)}{Y_{c,b}(Z)-Y_{c,u}(Z)},
\end{equation}

\begin{equation}\label{eq:normalization_h}
I(c,Z,h)=\frac{h-h_{min}(c,Z)}{h_{max}(c,Z)-h_{min}(c,Z)}.
\end{equation}

While the formulation described above provides a complete description of the flame structure in phase space, its application to turbulent combustion requires a statistical treatment of the governing equations. In the \gls{les} framework, the large-scale features of the flow are resolved, while the effects of smaller sub-filter scales must be modelled. To this end, the laminar transport equations are subjected to a Favre-filtering operation, introducing sub-filter terms that account for turbulent transport and the interaction between turbulence and chemistry at the sub-filter scale.

To obtain the transport equations for the Favre-filtered controlling variables, we follow the same procedure commonly adopted in previous tabulated-chemistry formulations for turbulent combustion~\cite{Dinesh2013,Donini2017,Almutairi2023}. Applying Favre filtering to the governing equations given by Eq.~\eqref{eq:eq_Y_control_general} yields the \gls{les} filtered transport equations for $\widetilde{Y}_c$, $\widetilde{Z}$, and $\widetilde{h}$:
\vspace{-2pt}
\begin{multline}\label{eq:filt_Yc}
\frac{\partial (\overline{\rho} \, \widetilde{Y}_c)}{\partial t} 
+ \nabla \cdot (\overline{\rho} \, \widetilde{\mathbf{u}} \, \widetilde{Y}_c) = \overline{\dot{\omega}}_{Y_c} \\
+ \nabla \cdot \Biggl(
    \overline{\rho} \left[\widetilde{\Gamma}_{Y_c,Y_c} + \frac{\nu_{t}}{\mathit{Sc}_{t}}\right] \nabla \widetilde{Y}_c
    + \overline{\rho} \, \widetilde{\Gamma}_{Y_c,Z} \, \nabla \widetilde{Z}
\Biggr)
\end{multline}
\vspace{-2pt}
\begin{multline}\label{eq:filt_Z}
\frac{\partial (\overline{\rho} \, \widetilde{Z})}{\partial t} 
+ \nabla \cdot (\overline{\rho} \, \widetilde{\mathbf{u}} \, \widetilde{Z}) = \\
 \nabla \cdot \Biggl(
    \overline{\rho} \left[\widetilde{\Gamma}_{Z,Z} + \frac{\nu_{t}}{\mathit{Sc}_{t}}\right] \nabla \widetilde{Y}_c
    + \overline{\rho} \, \widetilde{\Gamma}_{Z,Z} \, \nabla \widetilde{Z}
\Biggr)
\end{multline}
\vspace{-2pt}
\begin{multline}\label{eq:filt_h}
\frac{\partial (\overline{\rho} \, \widetilde{h})}{\partial t} 
+ \nabla \cdot (\overline{\rho} \, \widetilde{\mathbf{u}} \, \widetilde{h}) = \\
 \nabla \cdot \Biggl(
    \overline{\rho} \left[\widetilde{\Gamma}_{h,h} + \frac{\nu_{t}}{\mathit{Sc}_{t}}\right] \nabla \widetilde{h}
    + \overline{\rho} \, \widetilde{\Gamma}_{h,h} \, \nabla \widetilde{h}
\Biggr)
\end{multline}

\noindent where a gradient flux hypothesis has been applied to model the sub-filter fluxes and $\mathit{Sc}_{t}$ denotes the turbulent Schmidt number. Additional governing equations can be obtained for higher moments, such as the Favre sub-filter variances of progress variable variance $Y_{c,v} := \widetilde{Y_c^2} - \widetilde{Y}_c^2$ and mixture fraction variance $Z_v := \widetilde{Z^2} - \widetilde{Z}^2$.
The transport equations for $Y_{c,v}$ and $Z_v$ are given as follows:
\begin{multline}\label{eq:var_Yc}
\frac{\partial(\overline{\rho}\,Y_{c,v})}{\partial t}+\nabla\cdot(\overline{\rho}\,\widetilde{\mathbf{u}}\,Y_{c,v}) =
\nabla\cdot\left(\overline{\rho}\,\left[\widetilde{\Gamma}_{Y_{c},Y_{c}}+\frac{\nu_{t}}{\mathit{Sc}_{t}}\right]\nabla Y_{c,v}\right) \\
+2\overline{\rho}\,\frac{\nu_{t}}{\mathit{Sc}_{t}}\nabla\widetilde{Y}_{c}\cdot\nabla\widetilde{Y}_{c} \\
-\overline{\rho}\,C_{Y_{c}}\frac{\nu_{t}}{\mathit{Sc}_{t}}\frac{Y_{c,v}}{\Delta^{2}} +2\,(\overline{Y_{c}\,\dot{\omega}}_{Y_{c}}-\widetilde{Y}_{c}\,\overline{\dot{\omega}}_{Y_{c}}),
\end{multline}
\vspace{-2pt}
\begin{multline}\label{eq:var_Z}
\frac{\partial (\overline{\rho} \, Z_v)}{\partial t} + 
\nabla \cdot (\overline{\rho} \, \widetilde{\mathbf{u}} \, Z_v) = \nabla \cdot \left(\overline{\rho} \, \left[\widetilde{\Gamma}_{Z,Z} + \frac{\nu_{t}}{Sc_{t}} \right] \nabla Z_v\right) \\ + 2 \overline{\rho} \, \frac{\nu_{t}}{Sc_{t}} \nabla \widetilde{Z} \cdot \nabla \widetilde{Z}
 - \overline{\rho} \, C_Z \frac{\nu_{t}}{Sc_{t}} \frac{Z_v}{\Delta^2},
\end{multline}
\noindent where scalar dissipation rate of the variances is modulated with constants $C_{Y_c}$ and $C_Z$. Enthalpy variances are not considered.

The coefficients in these equations are obtained by integrating the manifold according to a presumed \gls{jpdf} in $(c,Z,I)$, where $I$ is a normalised enthalpy~\cite{Govert2018}. Since no sub-filter variance is considered for the enthalpy, the distribution in $I$ is represented by a Dirac delta centred at the filtered value $\widetilde{I}$, so that the filtered vector for the thermochemical state is given by:

\begin{equation}\label{eq:integration_jPDF}
\widetilde{\boldsymbol{\Psi}}
= \int_0^1\!\!\int_0^1
\boldsymbol{\Psi}(c,Z,\widetilde{I})\,
P(c,Z;\widetilde{c},c_v,\widetilde{Z},Z_v)
\, dc\, dZ.
\end{equation}

Under the additional assumption of statistical independence between $c$ and $Z$, the \gls{jpdf} factorises as:
\begin{equation}
P(c,Z;\widetilde{c},c_v,\widetilde{Z},Z_v)
= P_c(c;\widetilde{c},c_v)\, P_Z(Z;\widetilde{Z},Z_v),
\end{equation}
where $\widetilde{c}$ and $c_v$ denote the filtered mean and sub-filter variance of the normalised progress variable.
In this work, beta PDFs are used to model both $P_c$ and $P_Z$. A recent study by Berger et al.~\cite{BergerUpcoming} examined this dependence in detail and proposed strategies to account for it. Assessing the validity of the statistical independence assumption is left for future work.

\section{Numerical framework}\label{sec:numerical_fwrk}

The \gls{dns} simulation corresponds to that of Berger et al.~\cite{Berger2022_dns}, to which the reader is referred for numerical details, and is used here for the assessment of the \gls{les} solutions. A skeletal mechanism with nine species and nineteen reactions for hydrogen–air combustion, proposed by Burke et al.~\cite{Burke2012}, is employed in both the \gls{dns} and \gls{les} simulations. For the \gls{dns}, constant Lewis numbers obtained in the burnt mixture from one-dimensional flame calculations are imposed~\cite{Berger2022_dns}, whereas for the \gls{les}, molecular diffusivity is modelled using the mixture-averaged diffusion formulation. Note that the codes use different numerical methods, which could influence the direct comparison between solutions.
One-dimensional flame results confirm that constant Lewis numbers and mixture-averaged transport yield nearly identical global flame parameters across the range of equivalence ratios encountered in the jet simulation (see supplementary material). Consequently, the choice of diffusion model should have a negligible influence on the 3D results, a conclusion further supported by the simulation results presented in this work.

For the \gls{les}, the governing equations for continuity, momentum, enthalpy, controlling variables\ef{,} and associated sub-filter variances given by Eqs.\eqref{eq:filt_Yc}-\eqref{eq:var_Z} are solved under the low-Mach-number assumption, with the mixture treated as an ideal gas.
Sub-filter fluxes are modelled using the gradient flux hypothesis, and the Vreman model~\cite{Vreman2004} is applied to define the sub-grid eddy viscosity, with a model constant of $C_k=0.1$.
In these simulations, the turbulent Schmidt number is set to $Sc_{t}=0.7$, while constants $C_{Y_c}$ and $C_Z$ for the dissipation of the variances of the progress variable and mixture fraction are set as in previous studies~\cite{Govert2018,Both2020}.
The full set of equations is solved using the code Alya \cite{Vazquez2016}, which employs a low-dissipation, single-stage numerical scheme for velocity-pressure coupling \cite{Both2020}.
The scalar transport equations for the turbulent controlling variables are solved with a second-order spatial discretisation scheme based on the \gls{asgs} method and using linear finite elements \cite{Castillo2014}.
All governing equations are advanced in time using an explicit third-order Runge-Kutta scheme~\cite{Both2020}.

\vspace{-8pt}
\subsection*{Combustion model}
\vspace{-2pt}

Laminar flame manifolds are generated by solving unstretched, adiabatic one-dimensional premixed flames with the Cantera solver V3.0~\cite{cantera}, using the mixture-averaged diffusion model. The Soret effect is included following the Chapman–Cowling formulation for mixture-averaged transport~\cite{Chapman1999MathematicalTheoryOfNonUniformGases,Paul1998AReevaluationOfTheMeansUsedToCalculateTransportProperties}, a treatment that has recently been assessed in the context of hydrogen flames~\cite{Schlup2017ValidationOfAMixtureAveraged,Zirwes2025AssessmentOfApproximateSoret}.

The manifolds are constructed in two stages. First, a set of adiabatic, isobaric, premixed one-dimensional hydrogen–air flames is computed for an unburnt-gas temperature of $T_u = 298\mathrm{K}$. Next, flamelets at lower enthalpy levels are obtained using a burner-stabilised configuration~\cite{Oijen2001ModellingOfComplexPremixedBurnerSystems}, with the burner temperature fixed at the wall temperature, $T_{\mathrm{wall}} = 298\mathrm{K}$. A sequence of such flamelets is then computed by prescribing the inlet mass flux as a fraction of the mass flux of the corresponding adiabatic flame at $T_u$, progressively reducing it to achieve increasing levels of heat loss to the burner.

Water vapour is used as the progress variable, \(Y_c = Y_{\mathrm{H_2O}}\). 
For $\widetilde{Y}_c$, the discretisation consists of 101 linearly spaced points from $0$ to $1$ in normalised progress variable space. For the filtered mixture fraction and enthalpy, the discretisations include 58 and 17 points, respectively.
The scaled variance variables are discretised with 11 points 
following a cubic law.

\vspace{10 pt}
\section{Case configuration}\label{sec:case_config}

The case corresponds to a planar turbulent premixed jet flame with a main injection of hydrogen–air mixture at an equivalence ratio of 0.4 at atmospheric pressure and ambient temperature. The main injection is surrounded by a hot co-flow with a composition corresponding to the burnt gases at equilibrium considering the main jet composition (therefore, the mixture fraction values for both the inlet and the co-flow are equal).  The mixture exhibits an effective Lewis number $Le_{\text{eff}} \approx 0.346$ (estimated as in \cite{Joulin1981LinearStability}), which is below its critical Lewis number $Le_{c} \approx 0.824$ (calculated according to Sivashinsky’s theory for weak thermal expansion \cite{Sivashinsky1977DiffusionalThermalTheoryOfCellularFlames}).
The Karlovitz number based on the unstretched 1D flame speed and thickness is relatively low in this case (estimated at $Ka \approx 15$ \cite{Berger2022_dns}), which amplifies thermodiffusive effects and supports the applicability of scale-separation-based approaches such as flamelet models. Under these conditions, the combined influence of species diffusion and thermal diffusion introduces a destabilising effect on the flame. As demonstrated by Berger et al.~\cite{Berger2022_dns} through comparisons between DNS flames with $Le \neq 1$ and $Le = 1$, thermodiffusive effects have a leading order effect on the flame dynamics.

A three-dimensional computational domain is employed, consisting of a main hydrogen–air injection channel of length \(H = 8\,\text{mm}\) aligned with the \ef{\(y\)}-direction and bounded by solid walls of finite thickness \(w_t\) in the \ef{$x$}-direction.
The channel ends in a rectangular volume, which serves as the computational domain in this study, with only a section of height \(H\) from the injection channel as part of the domain.
The remainder of the cross-stream extent is supplied by the hot co-flow, and the domain is periodic in the spanwise (\(z\)) direction. The main injection has a bulk velocity of \(24\,\text{m}\,\text{s}^{-1}\)\ef{,} while the co-flow is injected at \(3.6\,\text{m}\,\text{s}^{-1}\). Given that the separation between the plates is \(H = 8\) mm, the jet Reynolds number for the main injection is $Re=11000$.

The \gls{dns} \dm{reference} case uses a structured hexahedral mesh \dm{with a total number of elements of $(N_x,N_y,N_z)=(1472, 1792, 512)$ and constant spacing \ef{$\delta_{x,z}^{\text{DNS}} = 70\, \mu \text{m}$}}. Turbulent inlet conditions are mapped directly from an auxiliary channel simulation. No-slip and isothermal \(T_{\text{wall}} = 298\,\text{K}\) boundary conditions are applied at the walls. The wall thickness is set to \ef{$w_t = 3 \delta_{x,z}^{\text{DNS}} = 0.21 \text{ mm}$}.

The \gls{les} also uses a structured mesh composed of hexahedra, but the mesh is coarsened towards the outlet in the \ef{$x$}-direction and the lateral sides in the \ef{$y$}-direction, so that the aspect ratio of the elements is kept constant in the diagonal, using geometrical expansion with ratios of 1.037 and 1.005 in the crosswise and streamwise directions, respectively. Four meshes (M1–M4) evaluate resolution effects by scaling element sizes relative to the wall thickness ($w_t$). \dm{Note that the present study uses implicit filtering, which means that each mesh resolution also features different LES filter size}. From M1 to M4, inlet and spanwise resolutions are 0.5, 1, 1.5, and 2 points per $w_t$, respectively. Wall resolution is 1 point per $w_t$ for M1–M2 and 2 points for M3–M4. This refinement increases the total element count from 2 M (M1) to 9.5 M (M2), 23 M (M3), and 42 M (M4). \ef{The meshes have a maximum resolution at the injection of 6, 3, 2 and 1.5 times $\delta_{\text{DNS}}$ and coarsen up to approximately 9.4, 6.5, 5.6 and 5 times $\delta_{x,z}^{\text{DNS}}$ at the average flame tip. Additional details of the meshes are provided in the supplementary material.}

Turbulent inlet conditions for each \gls{les} are provided by a set of precursor channel flow simulations with streamwise \ef{($x$)} and spanwise ($z$) periodicity at the corresponding \gls{les} resolution. The conditions are set for the same bulk Reynolds number using a wall model~\cite{Owen2020}. Fully developed spatial solutions are mapped to the jet inlet. Jet flame simulations apply a wall model to the channel walls and slip conditions to lateral boundaries.

The simulations considered in this study are given in Table~\ref{tab:cases}. \dm{One simulation including heat loss ($T_{wall}=298$ K) and Soret is performed for each mesh (M1 to M4), while two additional simulations using mesh M4 \ef{were} also conducted to
evaluate independently the effects of heat loss and thermodiffusion, `LES M4 no \(D_k^T\)' and `LES M4 no \(T_{\text{wall}}\)', respectively. }
The objective of these additional simulations is to provide complementary insights into the overall influence of different physical processes within tabulated-chemistry approaches.

\ef{
\setlength{\tabcolsep}{2.5pt} 
\begin{table}[t]
\centering
\small
\begin{tabular}{lccc}
\hline
Case & $T_{wall}$ & $D_k^T$ & CV \\ \hline
LES M4               & 298 K     & ON  & $(\widetilde{Z}, \widetilde{Y}_c, \widetilde{h}, Z_v, Y_{c,v})$ \\
LES M4 no $D_k^T$    & 298 K     & OFF & $(\widetilde{Z}, \widetilde{Y}_c, \widetilde{h}, Z_v, Y_{c,v})$ \\
LES M4 no $T_{wall}$ & Adiabatic & ON  & $(\widetilde{Z}, \widetilde{Y}_c, Z_v, Y_{c,v})$ \\
LES M3               & 298 K     & ON  & $(\widetilde{Z}, \widetilde{Y}_c, \widetilde{h}, Z_v, Y_{c,v})$ \\
LES M2               & 298 K     & ON  & $(\widetilde{Z}, \widetilde{Y}_c, \widetilde{h}, Z_v, Y_{c,v})$ \\
LES M1               & 298 K     & ON  & $(\widetilde{Z}, \widetilde{Y}_c, \widetilde{h}, Z_v, Y_{c,v})$ \\ \hline
\end{tabular}
\captionof{table}{Table of LES simulations. The column CV stands for controlling variables. The mesh refines from M1 to M4. }
\label{tab:cases}
\end{table}
}

\section{Results and discussion}\label{sec:results}
%
%
\begin{figure}
\centerline{\includegraphics[width=\linewidth]{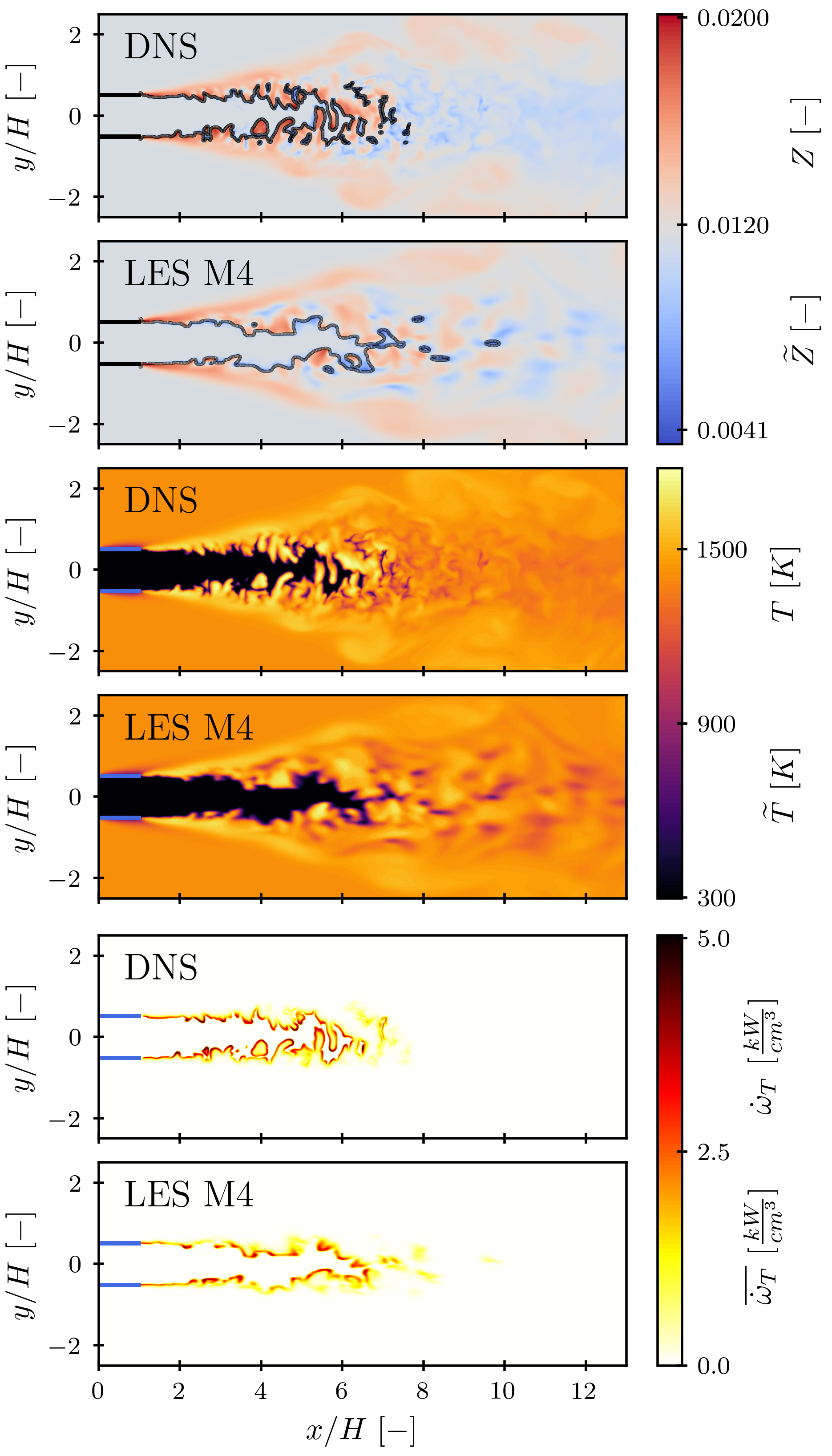}}
    \caption{Instantaneous snapshots of mixture fraction (first two rows), temperature (rows three and four), and heat release rate (rows five and six) for the DNS and the LES M4 (upper and lower plots of each pair, respectively). Level curves of scaled temperature $T_s$=0.6 from Eq.~\eqref{eq:norm_T} are included on the first two rows to aid the visualisation.
    \label{fig:instant}}
\end{figure}

This section provides a systematic comparison between the \gls{dns} reference solutions and six \gls{les} cases that employ different 
mesh resolutions and closures for thermodiffusive transport, and heat loss effects.
The analysis is performed to assess the ability of the proposed tabulated flamelet model to reproduce the fundamental characteristics of turbulent premixed hydrogen flames and its sensitivity to the mesh resolution (M1 to M4).

\subsection{Instantaneous fields}
\label{subsec:inst_fields}

In lean hydrogen flames, the low Lewis number promotes thermodiffusive instabilities that result in highly corrugated flame fronts. The combined action of this corrugation and turbulence increases the global consumption speed and flame surface area, while locally modifying the flame structure and reactivity ratio \cite{Berger2022_dns, Berger2022_partII}.
A key aspect of \gls{les} is the modelling of chemical source terms and transport properties required to predict the flame structure and dynamics. This becomes particularly relevant when thermodiffusive effects dominate, as unresolved sub-filter contributions may have some influence that is difficult to infer from the resolved scales. 
\dm{Consequently, assessing the impact of mesh resolution with their associated changes in filter size is essential to evaluate the accuracy and predictability of the proposed LES.}
In this section, a quantitative analysis based on the instantaneous fields is presented, with particular emphasis on the role of spatial resolution.

The instantaneous fields of mixture fraction $Z$, temperature $T$, and heat release rate $\dot{\omega}_T$ from the \gls{dns} along with the corresponding Favre-filtered \gls{les} quantities for the finest mesh (M4), are shown in  
Fig.~\ref{fig:instant}.
The visualisations illustrate the impact of thermodiffusive effects on flame morphology. To highlight the flame shape, the level curve corresponding to a scaled temperature $T_s=0.6$ is added on the subplots, where:
\begin{equation}\label{eq:norm_T}
T_s = \frac{T - T_{u}}{T^{1D}_{b} - T_u},
\end{equation}
\noindent and $T^{1D}_b = 1421.95$~K denotes the equilibrium temperature of the adiabatic one-dimensional flame at the inlet equivalence ratio. For the LES case, the same definition is retained, but the instantaneous $T$ \ef{is replaced} by instantaneous $\widetilde{T}$.

Two distinct regions are observed in physical space. The first corresponds to the primary reaction zone, characterised by an envelope of high heat release where alternating concave and convex fronts are present, forming finger-like structures, a characteristic feature of turbulence with thermodiffusive instabilities.
In regions where the flame front is concave (relative to the fresh gases),  thermodiffusive effects reduce the local mixture fraction, lowering temperatures and attenuating heat release. Conversely, convex regions exhibit local enrichment of the mixture, leading to super-adiabatic temperatures and intensified reaction rates. 
Further downstream of the primary reaction zone, pockets of both enriched and depleted mixture fraction emerge. However, depleted regions are predominant, leading to a heterogeneous, layered flow structure characterised by alternating zones of high and relatively low temperature.
The second region consists of two lateral branches that exhibit significant local enrichment and super-adiabatic temperatures, but with negligible heat release.
As illustrated in Fig. \ref{fig:instant}, these complex phenomena are successfully captured by the present turbulent combustion model.

\begin{figure}
\centerline{\includegraphics[width=\linewidth]{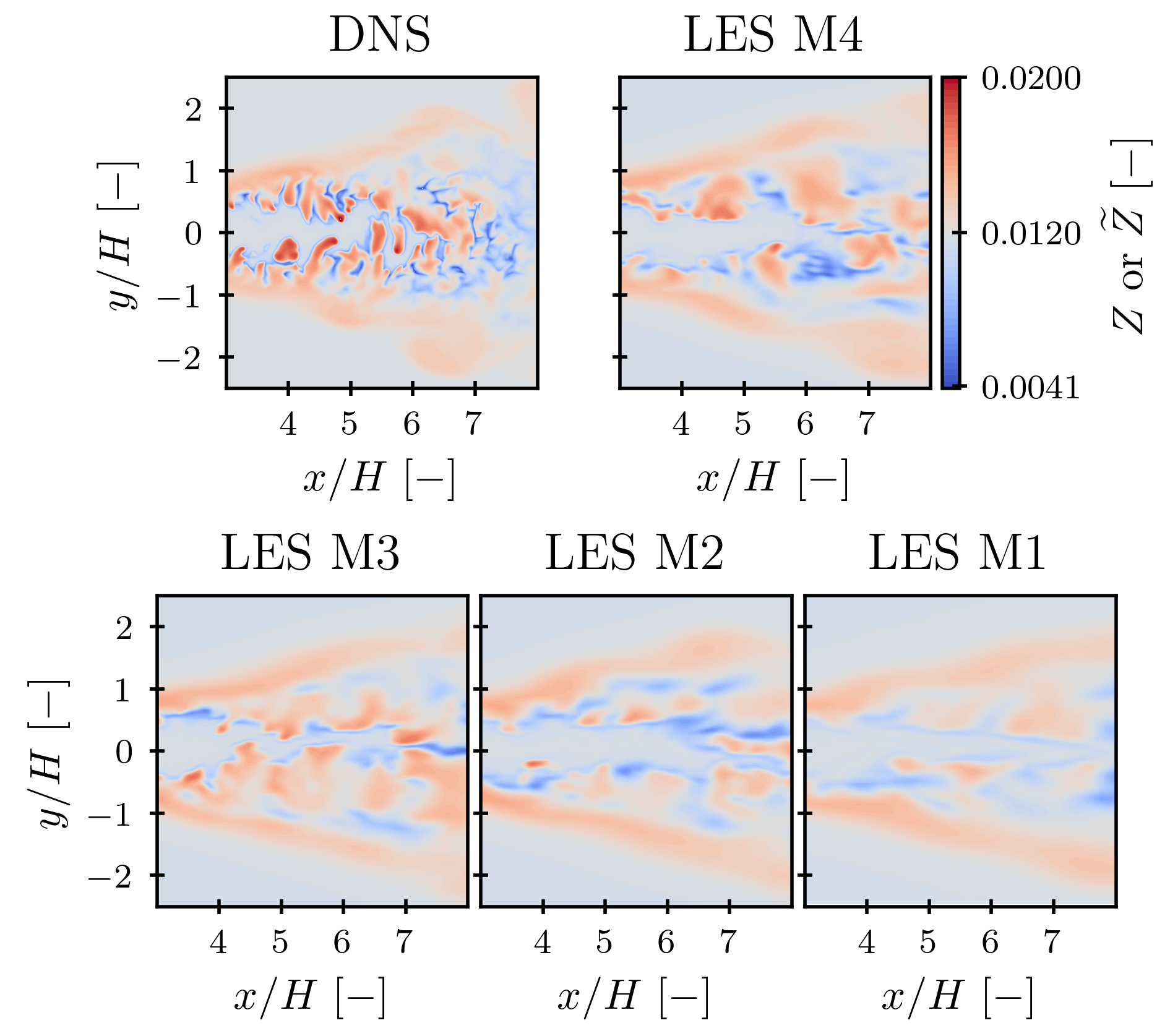}}
    \caption{Instantaneous snapshots of the $Z$ for the DNS and $\widetilde{Z}$ for LES for each mesh resolution.\label{fig:instant_window}}
\end{figure}

The influence of grid resolution on the first region is detailed in Fig.~\ref{fig:instant_window}. In the LES, due to the inherent resolution limits of the LES meshes and the use of low-order (second-order) numerical schemes, a significant portion of the fine-scale flame wrinkling remains unresolved and partially dissipated.  Since local variability in the mixture fraction is largely driven by flame curvature arising from both complex flame dynamics and thermodiffusive effects, this unresolved wrinkling directly results in attenuated fluctuations. Consequently, the LES mixture fraction field appears more uniform than the DNS. Furthermore, while the current model captures curvature-driven behaviour at resolved scales \cite{Fortes2024}, it lacks an explicit sub-filter formulation to recover the curvature fluctuations below the filter-size.
As a result, the correlation between the scalar fields in LES and DNS improves substantially as the mesh is refined from M1 to M4.

The instantaneous results show that the proposed approach can reproduce finger-like structures and super-adiabatic flame temperatures. While these features are highly unstable, they can influence the stationary flame structure. The statistically averaged behaviour of the jet is addressed in the next subsection.

%
%

\subsection{Mean fields}\label{subsec:mean_fields}

The temporally and spanwise-averaged fields (denoted by $\langle \cdot \rangle_{t,z}$) provide a more rigorous metric for assessing the ability of the model to predict the macroscopic flame topology. Averaging in $z$ is particularly useful, since it reduces the simulation time required to achieve statistically meaningful results. Furthermore, these mean representations allow for a systematic quantification of the primary modelling variables, specifically the sensitivity to mesh resolution and the physical impacts of Soret and heat loss. In this context, the focus shifts from individual wrinkling events to the spatial distribution of the reaction zone and the long-term stability of the lateral enriched branches.

To aid comparison, the contour corresponding to the scaled, filtered, and time- and spanwise-averaged temperature level $\langle T_s\rangle_{t,z}=0.6$ is superimposed on the plots, where:
\begin{equation}\label{eq:norm_T_filtered}
\langle T_s\rangle_{t,z} = \frac{\langle \widetilde{T} \rangle_{t,z} - T_{u}}{T^{1D}_{b} - T_u}.
\end{equation}
Using this quantity, the distance from the exit of the injector to the point where the iso-line $\langle T_s\rangle_{t,z}=0.6$ cuts the axis allows the definition of a flame length $h_T$.

\begin{figure}
\centerline{\includegraphics[width=\linewidth]{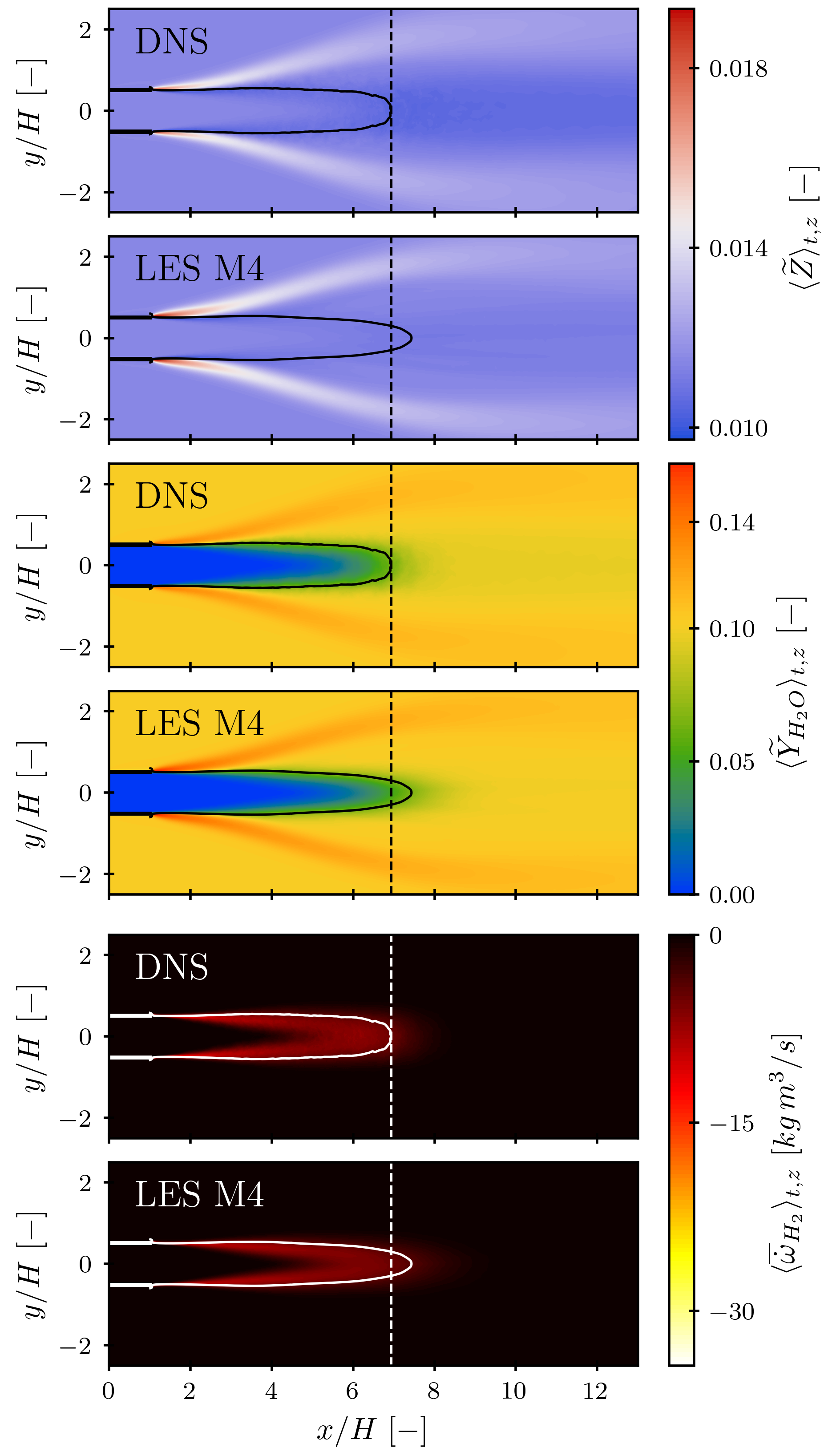}}
\caption{Mean fields of the mixture fraction  (first two rows), progress variable (rows three and four), and hydrogen source term (rows five and six) for the filtered DNS and the LES M4 (upper and lower plots of each pair, respectively). Level curves of scaled temperature $\langle T_s\rangle_{t,z}=0.6$ from Eq.~\eqref{eq:norm_T} are included to aid the visualisation.\label{fig:means_for_jpdf}}
\end{figure}

\begin{figure}

\centerline{\includegraphics[width=\linewidth]{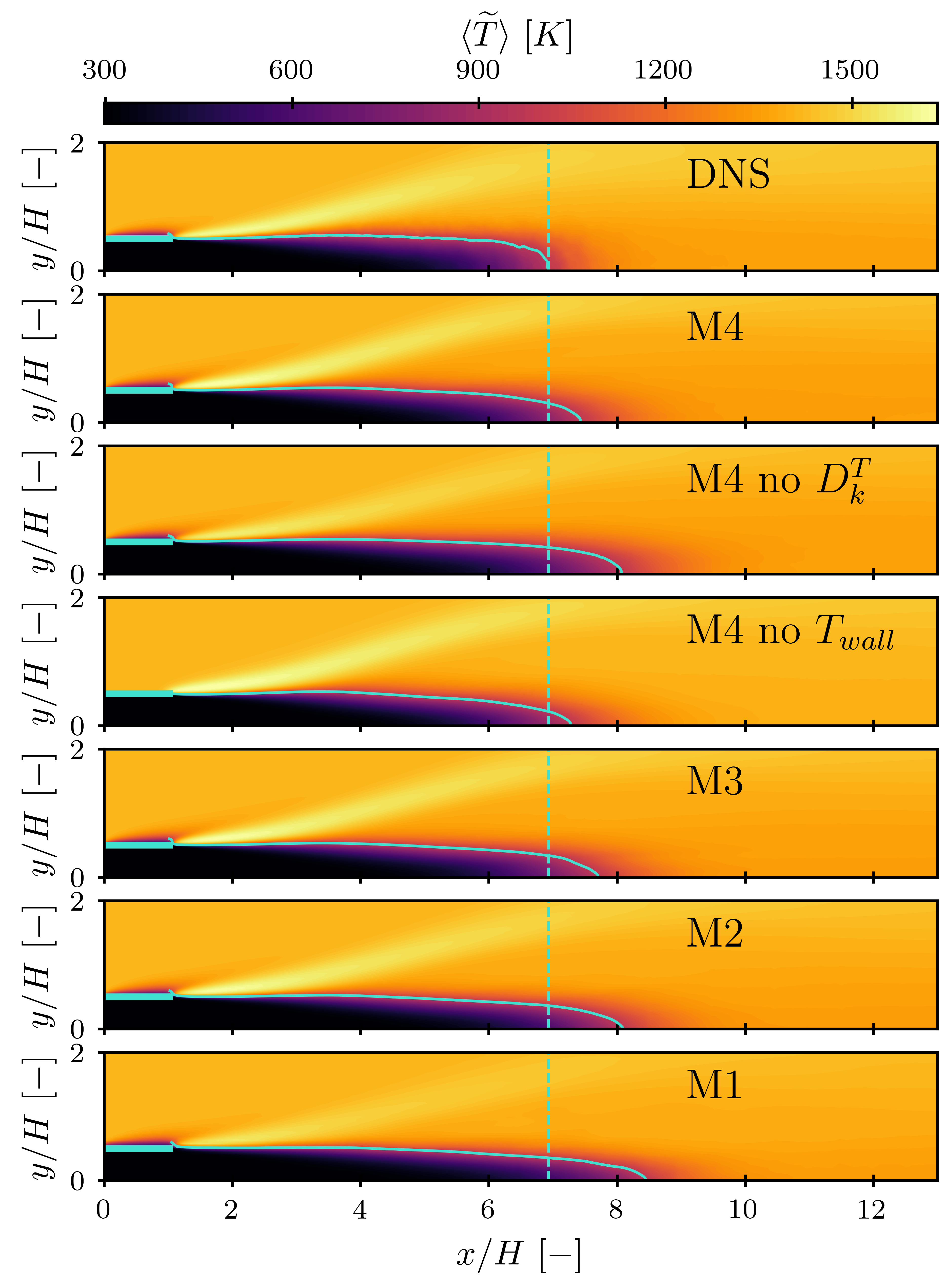}}
    \caption{$\langle \widetilde{T} \rangle_{t,z}$ field for the DNS (first panel), the LES with different resolutions and all physics (second to fifth panels), the LES without Soret effect (sixth panel) and the LES without heat losses (seventh panel).
    Isocontours of $sT = 0.6$ from Eq.\eqref{eq:norm_T}\ (cyan lines) and the DNS flame length $h_T$ are included for reference.\label{fig:mean_vs_res}}
\end{figure}

To establish a baseline for the high-resolution predictive capabilities, a side-by-side comparison of the DNS and LES M4 cases, focusing on the averaged controlling variables, $\langle \widetilde{Z} \rangle_{t,z}$ and $\langle  \widetilde{Y}_{H_2O}\rangle_{t,z}$, and the hydrogen source term, $\langle \overline{\dot{\omega}}_{H_2} \rangle_{t,z}$ is shown in Fig.~\ref{fig:means_for_jpdf}. This assessment is further extended in Fig.~\ref{fig:mean_vs_res}, which compares the averaged temperature fields, $\langle \widetilde{T} \rangle_{t,z}$, across all mesh resolutions and modelling configurations.

The \gls{les} results presented in Fig.~\ref{fig:means_for_jpdf} qualitatively reproduce the spatial flame structure observed in the \gls{dns}, including the two lateral branches. In these regions, both the progress variable and the temperature (see also Fig.~\ref{fig:mean_vs_res}) increase due to local enrichment of the mixture fraction induced by thermodiffusive effects. 
The \gls{les} slightly overestimates both the spreading angle of the lateral branches and the peak value of the mixture fraction and, consequently, those of temperature and progress variable, particularly near the wall edges. In addition, the \gls{les} predicts a slightly longer lateral branch than the \gls{dns}, although the difference remains small.

Downstream of the flame along the axis, the \gls{dns} exhibits a local reduction in the mixture fraction, resulting in sub-adiabatic temperatures. This decrease in mixture fraction is attributed to the conservation of the mixture fraction mass flow rate.
The \gls{les} qualitatively captures this behaviour, although the effect is slightly underestimated. 

Furthermore, the source term distribution indicates that the tabulated combustion model accurately reproduces the turbulent burning velocity, resulting in a flame length and shape that are in close agreement with the \gls{dns}. Minor differences are observed in the form of a slightly longer flame and sharper contours near the flame tip. These deviations could partly be attributed to the absence of a physical mechanism to account for sub-filter curvature effects in combination with all the introduced modelling decisions (variance equations, PDF integrations, numerical discretisation, eddy viscosity, among others). Overall, this represents a key outcome of the present study, demonstrating that the essential flame characteristics are captured despite the lack of explicit sub-grid curvature modelling in the proposed combustion framework.

%
%

%

Once the baseline flame topology has been established, it is necessary to separate the respective contributions of numerical resolution and physical modelling choices to the observed macroscopic behaviour. To this end, Fig.~\ref{fig:mean_vs_res} presents a 
\dm{dedicated} analysis of the averaged temperature field $\langle \widetilde{T} \rangle_{t,z}$, examining four mesh resolutions (M1–M4) as well as configurations in which individual physical modelling components are selectively disabled (M4 without $D_k^T$ and M4 without $T_{wall}$).

The \gls{les} results show that all mesh resolutions (M1–M4) qualitatively reproduce the spatial flame structure observed in the \gls{dns}, including the two lateral branches where temperature increases as a result of local mixture fraction enrichment induced by thermodiffusive effects.
In addition, the temperature field shows (qualitatively) little sensitivity to resolution, whereas the predicted flame height systematically improves and approaches the DNS value as mesh resolution increases. This is consistent with the improvement in capturing the source term (and consequently the consumption speed) when the resolution increases. This aspect is further clarified by the upcoming analysis of the influence of mesh resolution and thermodiffusion in the distribution of hydrogen's source term in the domain.
Finally, as the mesh resolution increases and a larger fraction of the turbulence–chemistry interaction is resolved, the flame tip evolves from a sharper to a more rounded shape, with the higher-resolution \gls{les} cases yielding flame geometries that are more consistent with those observed in the \gls{dns}.

Concerning the physical modelling, suppressing the Soret effect (M4 without $D_k^T$) results in a longer flame and lower temperatures in the lateral branches. This behaviour is consistent with the dual role of thermal diffusion in lean hydrogen flames. While the Soret effect typically decreases the one-dimensional laminar flame speed by approximately 5–10\%, it simultaneously enhances the diffusivity of light species. In lean hydrogen flames, this preferential transport amplifies thermodiffusive instabilities, thereby increasing the global consumption speed. As a result, simulations that account for the Soret effect exhibit shorter flames than those in which it is neglected \cite{Zhou2017EffectOfSoretDiffusion, Howarth2024ThermalDiffusion}. The lower temperatures observed in the lateral branches when thermal diffusion is omitted arise from the absence of preferential migration of light species, particularly $\mathrm{H}$ and $\mathrm{H}_2$, towards regions of high curvature or elevated temperature. This suppresses local mixture enrichment within the lateral cells, leading to lower temperatures and less reactive branch structures \cite{Zhou2017EffectOfSoretDiffusion, Howarth2024ThermalDiffusion}. Overall, these results highlight the importance of accounting for the Soret effect in hydrogen flame modelling and demonstrate that the present \gls{tc} approach successfully captures its dominant physical mechanisms.

\begin{figure}
 \centerline{\includegraphics[width=\linewidth]{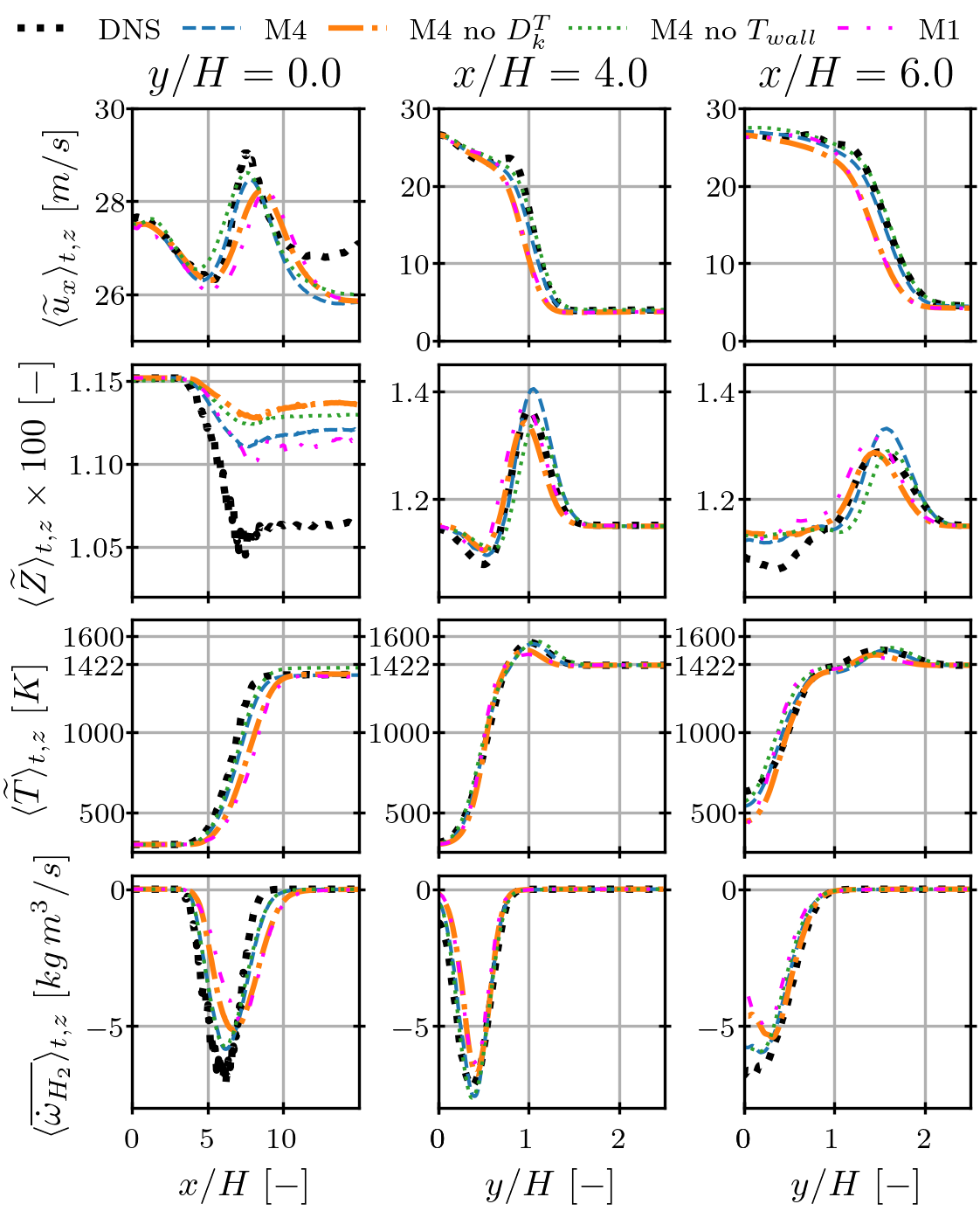}}
    \caption{\ef{Temporally and spatially averaged plots of $\langle \widetilde{u}_x \rangle_{t,z}$ (row 1), $\langle \widetilde{Z} \rangle_{t,z}$ (row 2), $\langle \widetilde{T} \rangle_{t,z}$ (row 3) and $\langle \overline{\dot{\omega}}_{H_2} \rangle_{t,z}$ (row 4) profiles along $y/H = 0$ (column 1), $x/H=4$ (column 2) and $x/H=6$ (column 3) for the DNS (dotted black lines), LES M4 (dashed blue lines), LES M4 no $D_k^T$ (dash-dot orange lines), LES M4 no $T_{wall}$ (dotted green lines) and LES M1 (dash-dot-dot pink lines).}
    \label{fig:profiles_axis}}
\end{figure}

When heat loss to the burner walls  is introduced, a new coordinate is included in the manifold and new terms are added into the governing equations for $Z$, $Y_c$ and $h$. Including these effects produces a slight improvement of the mixture fraction reduction observed downstream of the flame. Consequently, this region yields downstream temperatures closer to those of the co-flow. It should be noted, however, that the temperature difference remains limited to approximately 30~K (this point is visualised in more detail in Fig.~\ref{fig:profiles_axis}). 
The magnitude of this discrepancy remains relatively small, suggesting that the addition of the enthalpy coordinate in the manifold, primarily associated with wall heat losses, shows a limited impact on the LES solution.

\begin{figure*}
\centerline{\includegraphics[width=\linewidth]{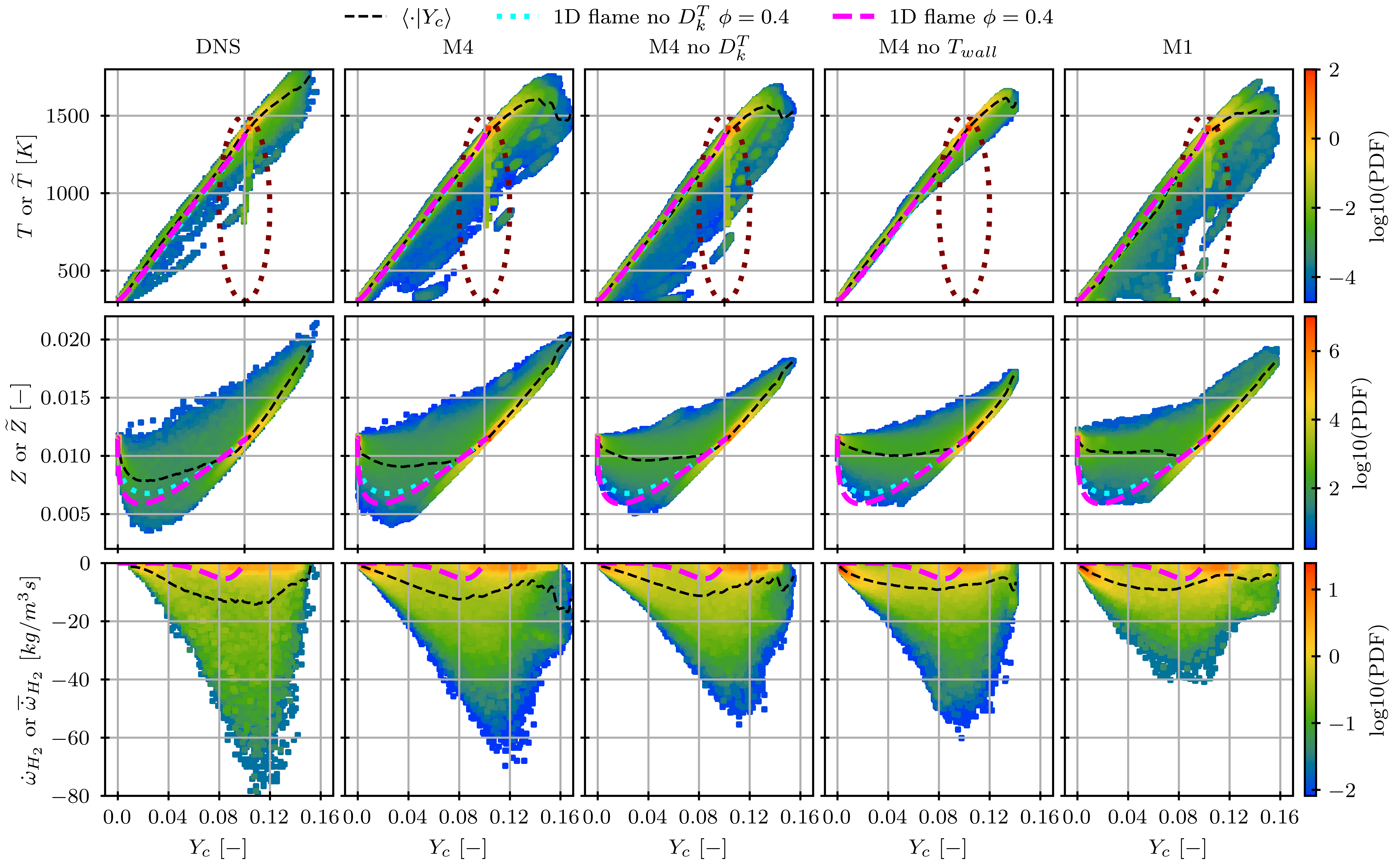}}
    \caption{\gls{jpdf} of temperature (first row), hydrogen source term (second row) and mixture fraction (third row) with respect to the progress variable for the DNS (first column), LES with resolutions M4 and M1 (columns two and three), LES without Soret effect (column four) and LES without wall heat losses (column five). A maroon ellipse is added to the first row to aid visualisation and highlight the region affected by heat losses.\label{fig:jPDFs}}
\end{figure*}

To move beyond qualitative comparisons and assess the thermochemical state of the mixture, the statistical distributions of the relevant scalars are examined. Figure~\ref{fig:jPDFs} shows the joint probability density functions (jPDFs) for the \gls{dns}, and the LES for case M4, including those without $D_k^T$, and without $T_{wall}$. This comparison enables an evaluation of whether the \gls{les} solutions follow the appropriate thermochemical manifold under the various modelling assumptions.
Each row shows the \gls{jpdf} of temperature, mixture fraction, and hydrogen source term as a function of the progress variable. While for the DNS instantaneous values are used, for the \gls{les}, these fields are represented using their resolved counterparts.
We have checked that filtering the DNS data for the comparison does not qualitatively alter the jPDF (not shown), and since it should be done for different filter sizes, we opt for just using the DNS value.
The same representation for the four mesh resolutions, including all physics, is available in the supplementary material.

%
%

For the temperature field, the scatter points follow the unstretched one-dimensional flame up to a progress variable of approximately 0.1, corresponding to the equilibrium value of the one-dimensional flame. Beyond this point, super-adiabatic temperatures arise from mixture fraction enrichment driven by curvature effects and interactions with the hot co-flow (lateral branches). As highlighted by the maroon ellipse in the figure, the simulations with wall heat losses exhibit an additional secondary branch of decreasing temperatures near the same progress-variable value, an effect absent in the adiabatic \gls{les} (column four). It is highlighted that when either coarsening the mesh or disabling the Soret effect, the jPDF extends over a wider region.
However, in all cases, including the adiabatic configuration, where flame spreading is reduced, the conditionally averaged temperature overlaps with the unstretched flame temperature over the range in which the latter is defined.

Both the hydrogen source term and the mixture fraction show clear improvements with increasing mesh resolution (cf. M1 versus M4), with the finer mesh leading to increased hydrogen consumption and the recovery of additional curvature-related features. Owing to thermodiffusive effects at the flame front, the mixture fraction shows noticeable scatter away from the one-dimensional flame manifold. However, the conditional mean remains close to the one-dimensional reference for both the \gls{dns} and the highest-resolution LES M4 case. A reduction in the spread of the mixture-fraction distribution is observed when either the Soret effect is disabled or the enthalpy coordinate is suppressed, indicating a diminished influence of thermodiffusive transport under these conditions.

Among the three quantities, the hydrogen source term is the most sensitive. It shows substantial improvement with mesh refinement, while the removal of either physical effect (Soret or enthalpy coordinate with heat loss) leads to a reduction in its magnitude relative to both the \gls{dns} and to LES M4. The conditional average of the source term differs significantly from that of the one-dimensional flame due to the strong sensitivity of the reaction rate to the combined effects of mixture fraction and progress variable.

Finally, to establish a more quantitative analysis, the axial and transverse profiles of key flow and scalar quantities are shown in Fig.~\ref{fig:profiles_axis}. \ef{The analysis focuses on the temporally and spanwise-averaged streamwise velocity $\langle \widetilde{u}_x \rangle_{t,z}$, mixture fraction $\langle \widetilde{Z} \rangle_{t,z}$, temperature $\langle \widetilde{T} \rangle_{t,z}$, and hydrogen reaction rate $\langle \overline{\dot{\omega}}_{H_2} \rangle_{t,z}$, evaluated along the centreline ($y/H=0$) and at two downstream transverse locations ($x/H=4$ and $6$).}

Overall, the finest-resolution case (LES M4) exhibits the closest agreement with the \gls{dns} across all examined quantities. The streamwise velocity profiles are well reproduced by all simulations, with the \gls{les} capturing the characteristic multi-peak structure. Near the flame base, the \gls{dns} shows an initial decrease in velocity up to approximately \ef{$x/H=5$}, associated with jet spreading. This behaviour is consistently recovered by all \gls{les} cases. Further downstream, the subsequent increase in velocity is linked to the local rise in temperature resulting from intense heat release, which modifies the density field and, through mass conservation, accelerates the flow.
Consistent with the preceding velocity decrease, the \gls{les} predictions accurately capture the axial locations of the temperature and heat-release peaks. Both simulations predict the maximum velocity\ef{,} and hence, the flame length at nearly the same axial position. 
The jet flame exhibits a progressive redistribution of momentum from the jet core to the surrounding flow, resulting in a gradual decrease in axial velocity and a corresponding reduction in heat release.
In the transverse direction, the velocity profiles decay smoothly until reaching the region affected by the lateral temperature branches induced by the co-flow interaction, beyond which the velocity asymptotically approaches the co-flow value.

Moreover, all simulations reproduce the sharp increase in temperature across the reaction zone, as evidenced by the axial profiles. It is noted, however, that the LES M4 case without the enthalpy coordinate exhibits a higher peak temperature, consistent with the earlier analysis of the mean temperature fields, with values approaching the adiabatic one-dimensional flame temperature ($T_b^{1D}=1421.95$~K, indicated on the temperature axis for reference), although the quantitative differences remain small from a practical standpoint. With respect to the mixture-fraction profile, the \gls{dns} shows a pronounced axial reduction associated with mixture-fraction conservation. This behaviour is partially captured by LES M4, although it underpredicts the final value as pointed out in the previous section.
In contrast, disabling either the Soret effect or the heat loss reduces the mixture fraction variation across the flame front. Changes in the mixture fraction within the lateral branches are also consistently captured across all cases. These reductions in mixture fraction correspond to the observed variations in the hydrogen reaction rate, of which all simulations reproduce the location and shape of the reaction-rate peak with good accuracy. Overall, the \gls{les} results are in good agreement with the DNS, with the largest deviations found in LES M1, highlighting the influence of mesh resolution. The comparisons emphasise that mesh resolution, thermal diffusion, and enthalpy effects each play a specific role in accurately reproducing the DNS flame structure and reaction-rate dynamics. The same figure, including all mesh resolutions, is provided in the supplementary material.

\subsection{Macroscopic parameters of the flame}
\label{subsec:flame_macroparam}

Whereas the previous sections assessed the accuracy of local thermodiffusive properties, the present section characterises the global flame behaviour using statistically averaged quantities, with a focus on the model's ability to predict integrated metrics that are directly relevant to practical applications.

We first evaluate the fuel flux as a function of height \ef{($\mathcal{F}(x)$)}, defined as the temporal and spanwise average of the $\mathrm{H_2}$ mass flux, followed by the integration over the remaining spatial directions:

\ef{
\begin{equation}\label{eq:axial_fuel_consumption}
    \mathcal{F}(x) = \frac{1}{\rho_u U H Y_{\mathrm{H}_2,u}} \int \, \langle \overline{ \rho u_x Y_{H_2} }
    \rangle_{t,z} (x, y) \, dy,
\end{equation}
}

\noindent where $\rho_u$ is the unburnt gas density, $U$ is the bulk inlet velocity of the jet, $H$ is the slot width, and $Y_{\mathrm{H}_2,u}$ is the hydrogen mass fraction in the unburnt gases.

\begin{figure}
\includegraphics[width=\linewidth]{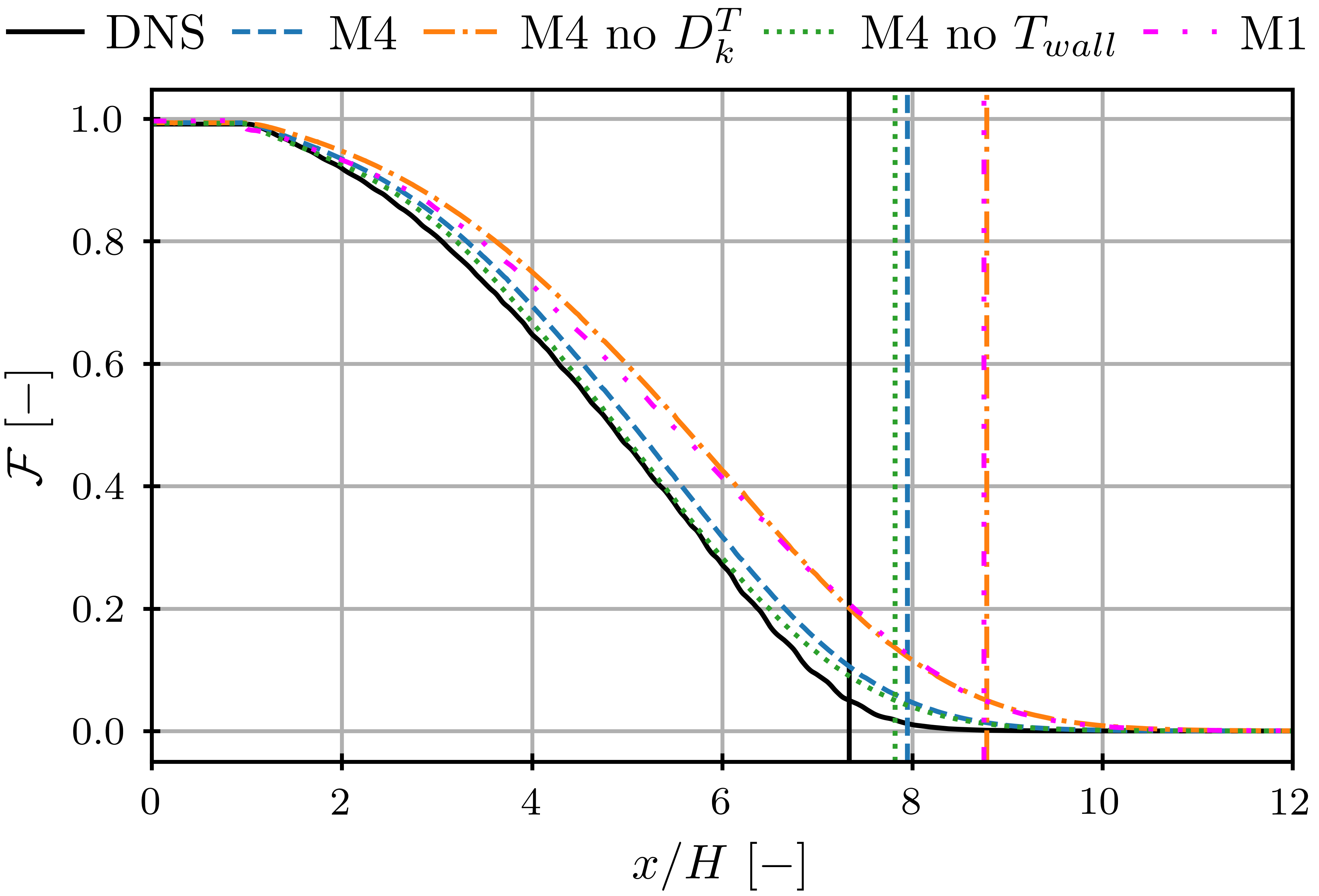}
\caption{Fuel consumption for the DNS and the LES study cases. The flame height $h_{\mathcal{F}}$ is included with vertical lines for the different cases.
\label{fig:fuel_cosump}}
\end{figure}

The axial evolution of fuel flux is shown in Fig.~\ref{fig:fuel_cosump} for the \gls{dns} and all \gls{les} modelling strategies, providing a clear illustration of the influence of transport modelling and mesh resolution on the global flame length. The plot shows a good correlation between the \gls{dns} and LES M4 results. Adding the heat loss through a
new manifold coordinate for the enthalpy leads to a modest reduction in the fuel-consumption rate, though this effect may be considered small in this case. In contrast, suppressing the Soret effect has a greater impact, as the associated reduction in local burning rates results \ef{in} a substantial increase in the predicted flame height. These results confirm that, for lean hydrogen mixtures, neglecting thermal diffusion systematically overpredicts the flame length.

In addition, the flame height, consumption speed, and flame topology, quantified through the flame surface perimeter, are examined. The flame height is first evaluated using the previously defined metric $h_T$, corresponding to the maximum \ef{$x$-coordinate} of a closed temperature isoline obtained from Eq.~\eqref{eq:norm_T_filtered} at $\langle sT\rangle = 0.6$. Furthermore, the definition of the fuel consumption given in Eq.~\eqref{eq:axial_fuel_consumption} enables the introduction of an alternative flame-height measure, $h_{\mathcal{F}}$, defined by a threshold criterion at $\mathcal{F}=0.05$ \cite{Berger2022_dns}.

The consumption speed, $\overline{s_c}$, defined as the averaged hydrogen-consumption rate, and the flame perimeter (instead of surface due to the averages in $z$), $l\left(\langle sT\rangle\right)$, where $\langle sT\rangle$ denotes the isoline of the averaged scaled temperature defined in Eq.~\eqref{eq:norm_T_filtered}, are evaluated along the corresponding mean temperature isoline as follows:

\begin{equation}\label{eq:average_macro_l}
    l\left(\langle sT\rangle \right)= \text{length of isoline at value $\left(\langle sT\rangle \right)$}.
\end{equation}

\begin{equation}\label{eq:average_macro_sC}
    \overline{s_c} \left(\langle sT\rangle \right) = -\frac{1}{\rho_u Y_{\mathrm{H}_2,u} l\left(\langle sT\rangle \right)} \int_A \langle \overline{\dot{\omega}_{\mathrm{H_2}}} \rangle_{t, z}\, dS,
\end{equation}

Table~\ref{tab:macroparameters} summarises the results for the different flame height definitions, consumption speed, and flame surface perimeter for the \gls{dns} and all \gls{les} cases.

%
%

\setlength{\tabcolsep}{3pt} 
\begin{table}[t]
\centering
\small
\begin{tabular}{lrrrr}
\toprule

%
%
Case & $h_{\mathcal{F}}/H$ & $h_{T}/H$ & $l/H$ & $\overline{s_{c}}/s_{L}^{0}$ \\
\midrule
DNS & 7.34 & 6.93 & 12.39 & 11.27 \\
LES M4 & 7.95 & 7.43 & 13.20 & 9.88 \\
LES M4 no $D_k^T$ & 8.78 & 8.07 & 14.46 & 8.67 \\
LES M4 no $T_{wall}$ & 7.82 & 7.28 & 12.75 & 10.04 \\
LES M3 & 8.10 & 7.71 & 13.69 & 8.84 \\
LES M2 & 8.46 & 8.10 & 14.38 & 8.71 \\
LES M1 & 8.75 & 8.45 & 15.02 & 7.76 \\

\bottomrule
\end{tabular}
\captionof{table}{\gls{les} macro-parameters. Results for the consumption speed are scaled with the unstretched laminar flame speed $s_L^0$ at $\phi=0.4$ with constant Lewis numbers for the DNS and mixture-averaged transport for the LES.}
\label{tab:macroparameters}
\end{table}

The flame height is generally comparable across all simulations. Increasing the mesh resolution leads to clear convergence towards the \gls{dns}, although this quantity is not particularly sensitive to resolution, and even coarser meshes provide reasonably accurate predictions. The \gls{les} effectively captures both consumption speed and flame-surface perimeter, with relative errors on the finest mesh restricted to 12.33\% and 6.54\%, respectively. For the flame height measures, the errors are also in the same ranges, with 8.31\% for $h_{\mathcal{F}}$, and 7.22\% for $h_{T}$. 
Consistent with previous observations, neglecting the Soret effect leads to a reduction in the consumption speed with a corresponding increase in the flame length, highlighting the importance of thermal diffusion even for turbulent flames simulated with \gls{les} where sub-filter effects play an important role. In contrast, the influence of heat loss remains modest in this configuration since the interaction with the walls is minor.
Overall, these findings demonstrate the robustness and reliability of the proposed tabulated flamelet model for predicting key flame characteristics using a simple turbulent chemistry interaction model based on presumed-shape PDF with beta functions in mixture fraction and progress variable. The results indicate that this modelling approach is sufficient for the present conditions, while more complex closures may be required to fully account for unresolved sub-filter effects on more complex scenarios.

\section{Conclusions and future work}\label{sec:conclusions}

In this study, a tabulated flamelet model accounting for differential and preferential diffusion is applied to large-eddy simulations of a planar turbulent premixed jet flame over a range of mesh resolutions and different physical features.  
The results are compared with those from the reference DNS of Berger et al.~\cite{Berger2022_dns} showing that the model accurately reproduces the global flame structure and its main characteristics, while providing insights into the role of unresolved sub-filter effects on local thermochemical behaviour.

The turbulent burning velocity and corresponding flame height are well predicted, as confirmed by the mean temperature fields. The interaction between the jet and the co-flow is also captured, including the formation of laterally enriched mixture fraction branches driven by differential diffusion, which lead to locally super-adiabatic temperatures. While the location of these branches is accurately reproduced, a slight overprediction of the mixture enrichment is also observed.

Limitations related to sub-filter curvature effects are identified but remain minor for the conditions considered. With increasing mesh resolution, a larger fraction of the turbulence–chemistry interaction is resolved, resulting in smoother flame-tip geometries and improved agreement with the rounded flame shape observed in the DNS. Quantitative comparisons of axial and transverse profiles show good overall agreement, with discrepancies decreasing systematically with mesh refinement. An assessment of global flame behaviour based on statistically averaged and integrated metrics shows good agreement with the DNS. The analysis confirms that the model accurately predicts flame height, consumption speed, and flame morphology, with limited sensitivity to mesh resolution and heat loss under the present conditions.

The analysis of thermal diffusion highlights the role of the Soret effect when modelling hydrogen flames, which enhances mixture reactivity, increases the consumption speed, and shortens the flame, whereas its omission leads to significant deviations in the predicted behaviour. Incorporating heat loss effects through an additional manifold coordinate provides only marginal improvement since no significant heat-loss effects are present in this problem, and thus the adiabatic formulation remains sufficiently accurate for this case.

Overall, despite unresolved sub-filter effects, the proposed turbulent combustion model captures the primary flame structure and its dominant characteristics, supporting its extension to more advanced sub-filter turbulent-chemistry interaction models and its application to more practical applications of industrial interest.

\section*{CrediT authorship contribution statement}

\textbf{Emiliano Manuel Fortes}: Formal analysis, Investigation, Data curation, Software, Visualisation, Writing – original draft. 
\textbf{Eduardo Javier Pérez-Sánchez}: Supervision, Formal analysis, Investigation, Software, Visualisation, Writing – original draft. 
\textbf{Temistocle Grenga}: Supervision, Methodology, Formal analysis, Writing – original draft. 
\textbf{Michael Gauding}: Formal analysis, Resources, Writing – original draft. 
\textbf{Heinz Pitsch}: Formal analysis, Resources, Writing – original draft.
\textbf{Daniel Mira}: Supervision, Methodology, Formal analysis, Funding acquisition, Project administration, Resources, Writing – original draft. 

\section*{Declaration of competing interest}

The authors declare that they have no known competing financial interests or personal relationships that could have appeared to influence the work reported in this paper.

\section*{Acknowledgments}

The research leading to these results has received funding from the European Union’s Horizon 2020 Programme under 
the Horizon Europe HyInHeat project  GA 101091456 
and H2AERO CPP2022-009921  both funded by MICIU/AEI/10.13039/501100011033 and by "European Union NextGenerationEU/PRTR". EMF acknowledges the predoctoral grant Joan Oró-FI (2023 FI-1 00680) funded by AGAUR from the Secretariat of Universities and Research of the Department of Research and Universities of the Generalitat de Catalunya and the ESF+. DM acknowledges the Grant RYC2021-034654 funded by MICIU/AEI/10.13039/501100011033 and by ‘‘European Union NextGenerationEU/PRTR’’. EJPS acknowledges his AI4S fellowship within the "Generación D" initiative by Red.es, Ministerio para la Transformación Digital y de la Función Pública, for talent attraction (C005/24-ED CV1), funded by NextGenerationEU through PRTR. The authors also acknowledge computational resources by the EuroHPC allocation EHPC-REG-2023R03-194 and RES allocation IM-2025-2-0035 provided by BSC.

\section*{Supplementary material}\label{sec:supplementary}

Supplementary material is submitted along with the manuscript.
This supplementary material provides technical details supporting the modelling choices and the mesh sensitivity analysis presented in the main text. First, a visualisation of each mesh resolution is provided. Second, the thermochemical properties of the 1D flames are analysed to justify the transition from the constant-Lewis-number approach used in the DNS to the mixture-averaged transport model employed in the Large-Eddy Simulations (LES) manifold. 
Third, an extended sensitivity analysis is provided, including joint Probability Density Functions (jPDFs) and axial profiles for all mesh resolutions, further quantifying the convergence of the LES toward the reference DNS data.

\subsection*{Mesh resolutions}

\begin{figure}
\centerline{\includegraphics[width=1.0\linewidth]{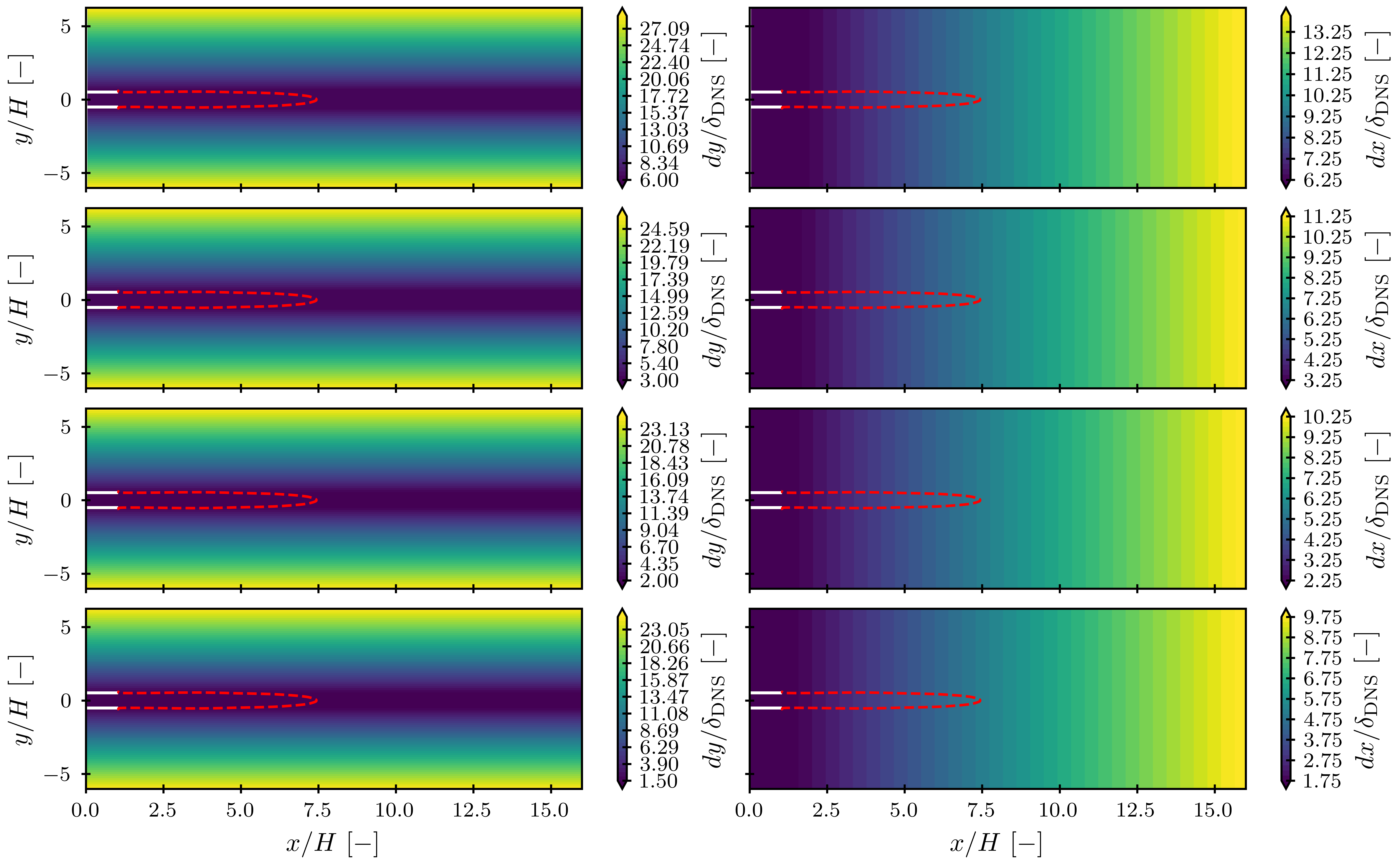}}
\caption{Contour plots of $dy / \delta_{x,z}^{\text{DNS}} $ (first column) and $dx / \delta_{\text{DNS}}$ (second column) for each mesh resolution M1 (first row), M2 (second row), M3 (third row) and M4 (fourth row). Dashed red lines indicate the flame perimeter for LES M4.
\label{fig:meshres}}
\end{figure}

Figure~\ref{fig:meshres} illustrates the mesh resolution in the directions of the $x$ and $y$ axes with respect to the dns spacing $\delta_{x,z}^{\text{DNS}} = $ 70 $\mu \text{m}$. The perimeter of scaled temperature defined in the main text for the simulation LES M4 is included for reference on the mesh resolution according and flame height. In the $z$ axes, the spacing is $dz / \delta_{x,z}^{\text{DNS}} = 6, 3, 2$ and $1.5$ for M1, M2, M3 and M4, respectively. 

\subsection{One-dimensional flames and transport validation}

Table~\ref{tab:main_cases} summarises the primary thermochemical properties of the one-dimensional (1D) flames at the jet injection conditions ($\phi=0.4$). To validate the consistency of the modelling framework, calculations were performed using three approaches: the mixture-averaged transport approximation (both with and without the Soret effect) and the constant Lewis number ($Le$) formulation adjusted according to the methodology of Berger et al. \cite{Berger2022_dns}. All 1D calculations were performed using the open-source software Cantera~\cite{cantera}.

The results demonstrate that the global flame macroparameters of laminar burning velocity ($s_L$) and thermal flame thickness ($l_F$) exhibit negligible deviations between the adjusted constant $Le$ and mixture-averaged models. This alignment provides a critical justification for benchmarking the current LES results against the DNS data. It ensures that any observed differences in the three-dimensional turbulent flame structure are attributable to sub-filter modelling and turbulence-chemistry interaction rather than fundamental discrepancies in the underlying laminar flame structure. Furthermore, comparing the mixture-averaged results with and without thermal diffusion shows that the Soret effect significantly influences the local consumption rate, necessitating its inclusion for an accurate representation of lean hydrogen flame chemistry.

\begin{table}
\centering
\begin{center}
\footnotesize 
\begin{tabular}{|c|c|c|c|c|}
\hline
Label & $D_k^T$ & $s_{L}$ [cm/s] & $l_{F}$ [$\mu$m] & $\tau$ [ms] \\
\hline
Mix-avg & On & 18.3 & 681.3 & 3.72 \\
Mix-avg no $D_k^T$ & Off & 19.2 & 665.8 & 3.47 \\
$Le$ cst & On & 18.0 & 698.7 & 3.91 \\
\hline
\end{tabular}
\end{center}
\caption{Flame properties of the main injection mixture at $\phi=$ 0.4, $T=$ 298K and $ p=$1 atm using mixture-averaged transport with and without thermal diffusion, and using constant Lewis numbers with thermal diffusion.}
\label{tab:main_cases}
\end{table}

\begin{figure}
\centerline{\includegraphics[width=0.5\linewidth]{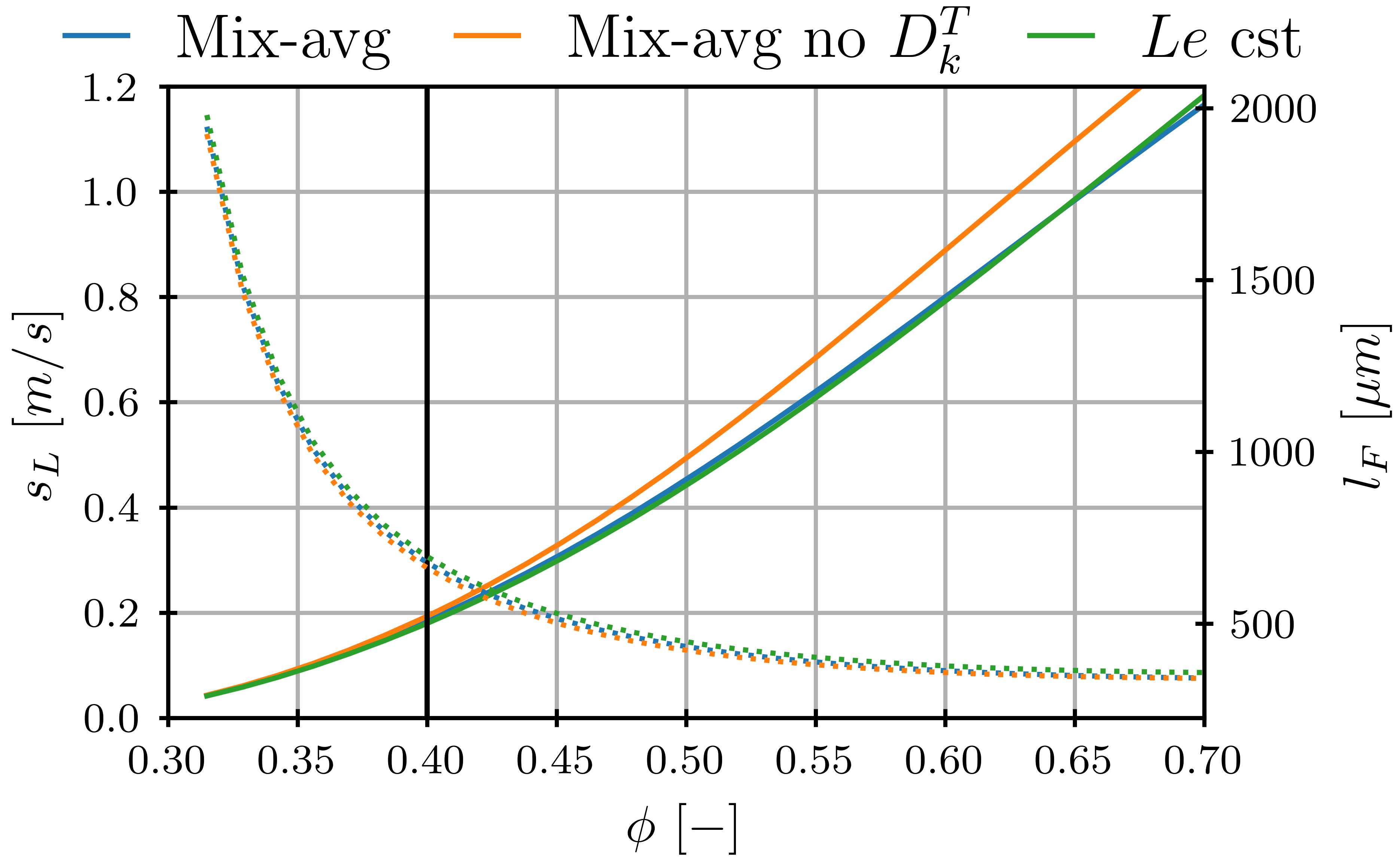}}
\caption{Comparison of laminar burning velocity $s_L$ (left $y$ axis) and thermal flame thickness $l_F$ (right $y$ axis) as a function of the equivalence ratio $\phi$ for hydrogen–air mixtures at $T_u = 298$~K and $p = 1$~atm. Solid lines denote $s_L$ for the mixture-averaged transport model with the Soret effect (blue lines), mixture-averaged transport without the Soret effect (orange lines), and the adjusted constant Lewis number formulation (green lines). The same colour code is used for the dotted lines in $l_F$.
\label{fig:1d_comparison}}
\end{figure}

To reinforce this validation, it is essential to verify the model agreement not only at the nominal injection equivalence ratio but across the entire range of $\phi$ values encountered in the turbulent jet. Fig.~\ref{fig:1d_comparison} illustrates the laminar burning velocity ($s_L$) and thermal flame thickness ($l_F$) as a function of the equivalence ratio ($\phi$). The results demonstrate negligible deviations between the adjusted constant Lewis number approach and the mixture-averaged transport model across the equivalence ratios relevant to this study. This agreement ensures a consistent and fair comparison between the DNS and LES frameworks. Additionally, results for the mixture-averaged transport model without thermal diffusion ($D_k^T = 0$) are included to highlight the magnitude of Soret effects on the flame properties.

\subsection{Extended mesh sensitivity analysis}

This section provides a comprehensive evaluation of the \dm{sensitivity of the LES results} to mesh resolution, ranging from the coarsest (M1) to the finest (M4) grids. The results are systematically compared against the reference \dm{DNS data} to quantify the convergence of both local thermochemical states and global flame parameters.

\begin{figure*}
\centerline{\includegraphics[width=\linewidth]{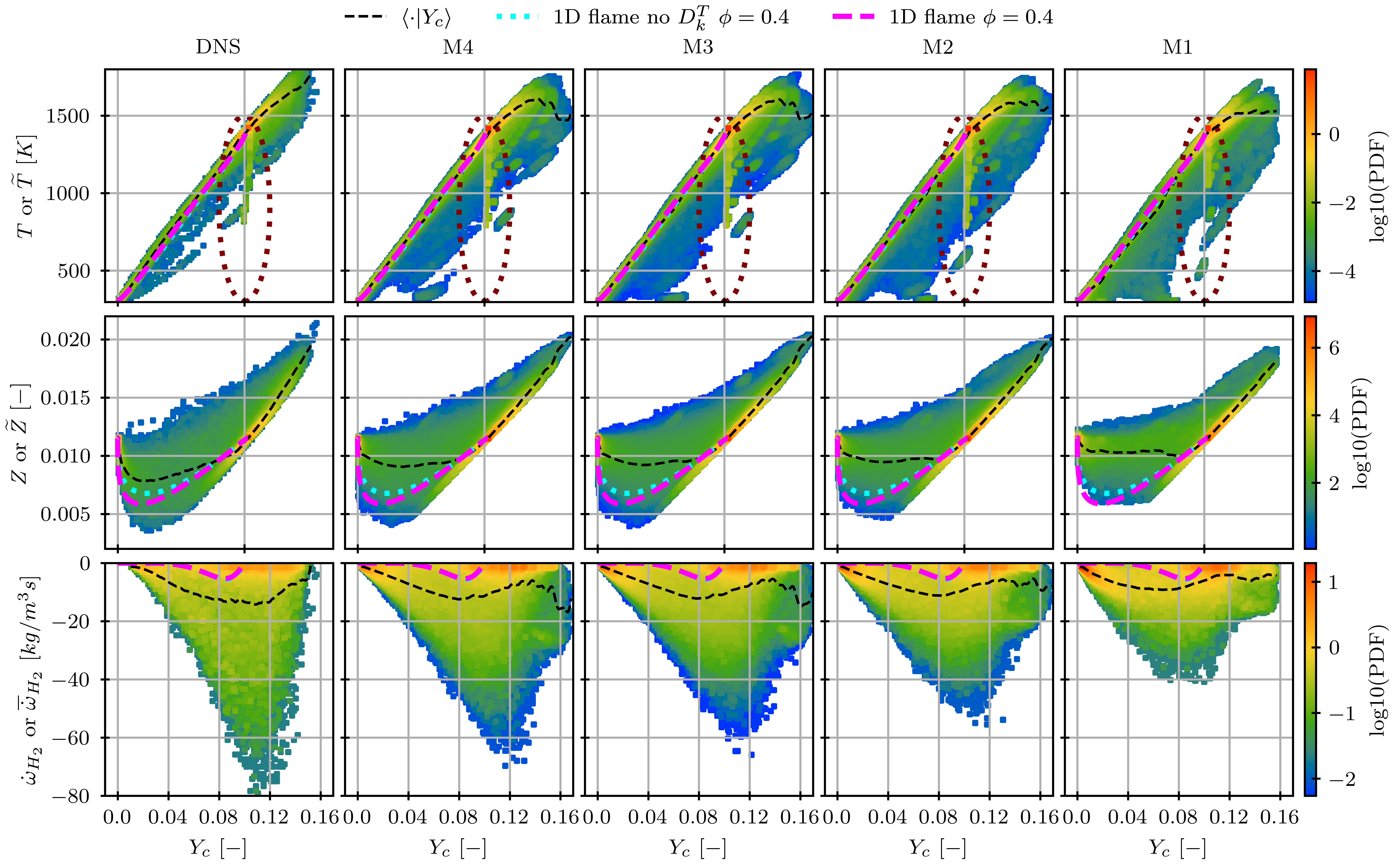}}
\caption{\gls{jpdf} of temperature (first row), hydrogen source term (second row) and mixture fraction (third row) with respect to the reactive progress variable for the DNS (first column), and the LES with resolutions M4, M1, M2 and M3 (columns two to five). \label{fig:sup_jPDFs}}
\end{figure*}

Figure~\ref{fig:sup_jPDFs} illustrates how mesh convergence systematically improves the jPDF distribution of the analysed quantities. As the mesh resolution \dm{is increased}, the mixture fraction field captures a broader scatter range, reflecting the increased resolution of local enrichment and depletion zones. A similar trend is observed for the source term, \dm{which remains the most grid-sensitive quantity, as finer meshes more accurately resolve high-reactivity peaks that are typically} smoothed out on coarser grids.
%
%

The impact of resolution on the mean flame structure is further quantified through axial and transverse profiles of velocity, temperature, mixture fraction, and hydrogen source term. While global features, such as the velocity decay, are captured even on coarser grids, the accurate prediction of the peak reaction rate, peak velocities, and the specific axial location of the flame tip requires \dm{finer resolution as shown by the comparative results from mesh M1 to M4.}
In contrast, the decay of the mixture fraction is less sensitive to mesh resolution, which suggests that further improvements \dm{would require more advanced sub-filter modelling.}

\begin{figure}
\centerline{\includegraphics[width=0.5\linewidth]{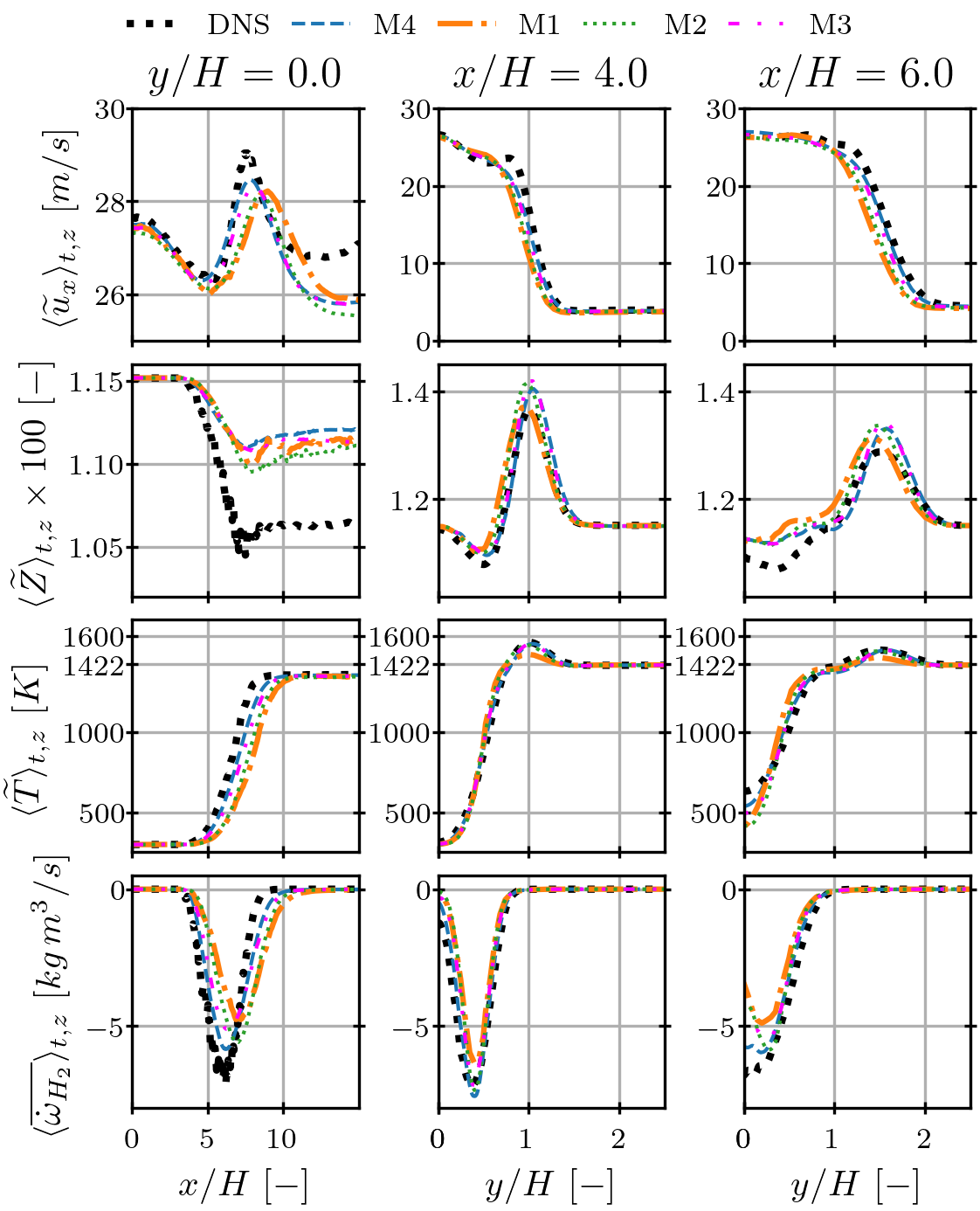}}
\caption{Temporally and spatially averaged plots of $\langle \widetilde{u}_x \rangle_{t,z}$ (row 1), $\langle \widetilde{Z} \rangle_{t,z}$ (row 2), $\langle \widetilde{T} \rangle_{t,z}$ (row 3) and $\langle \overline{\dot{\omega}}_{H_2} \rangle_{t,z}$ (row 4) profiles along $y/H = 0$ (column 1), $x/H=4$ (column 2) and $x/H=6$ (column 3) for the DNS (dotted black lines), LES M4 (dashed blue lines), LES M3 (dash-dot orange lines), LES M2 (dotted green lines) and LES M1 (dash-dot-dot pink lines).
\label{fig:sup_profiles_axis}}
\end{figure}

\dm{Finally, the influence of mesh resolution on fuel consumption is clearly evident, with the integrated consumption rate converging towards the DNS reference as the grid is refined. The small differences observed between the M3 and M4 cases indicate that statistical convergence and almost mesh independence for this global metric are effectively achieved at the higher resolutions.}

\begin{figure}
\centerline{\includegraphics[width=0.5\linewidth]{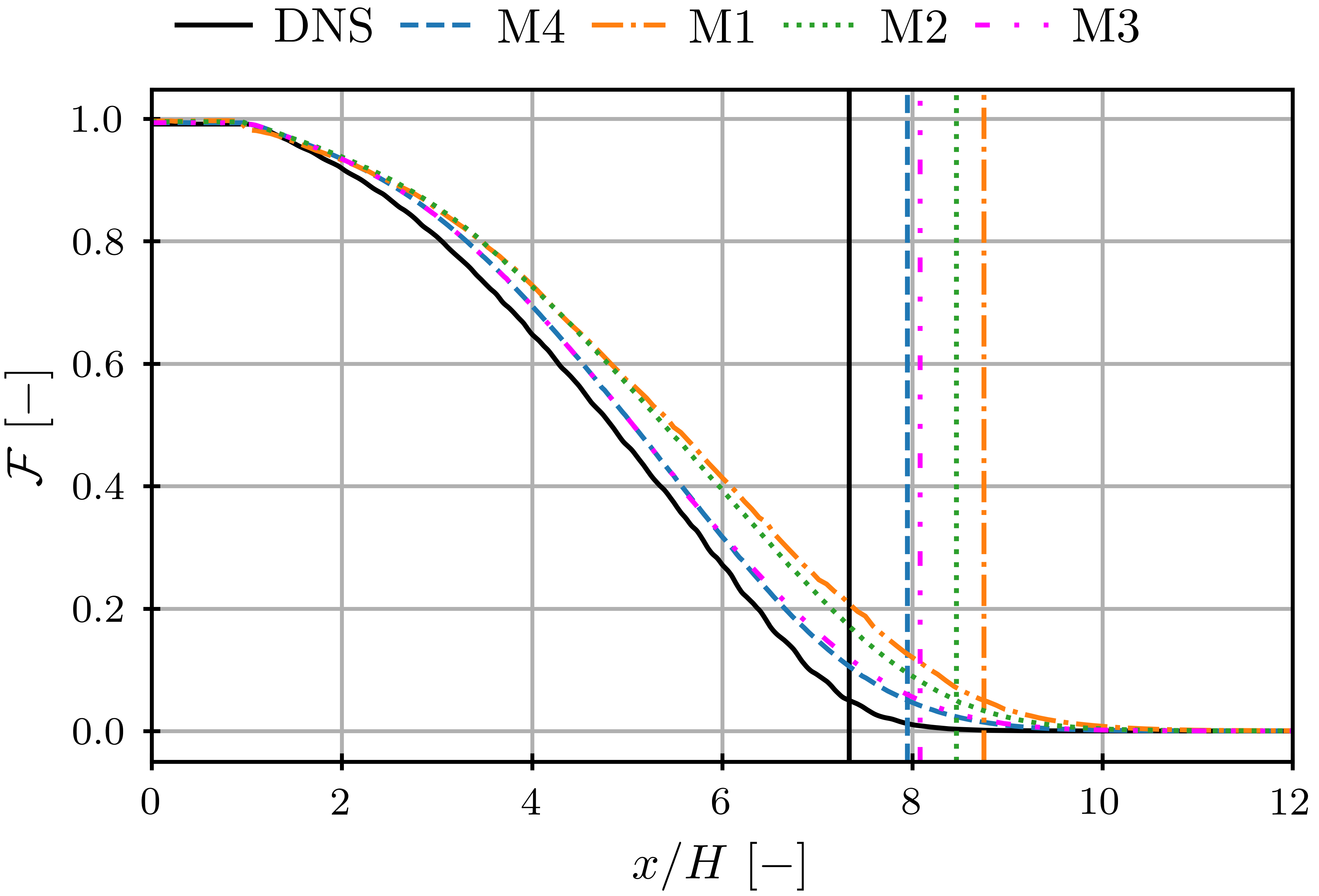}}
\caption{Fuel consumption for the DNS and the LES with all mesh resolutions. Flame height $h_{\mathcal{F}}$ is included with vertical lines.
\label{fig:sup_fuel_cosump}}
\end{figure}

\FloatBarrier

\bibliographystyle{cnf-num}
\bibliography{biblio}

\end{document}